\documentclass[12pt,a4paper,fleqn]{elsarticle}

\usepackage{lineno}
\modulolinenumbers[5]

\usepackage{euscript,amsmath,amssymb,color}
\usepackage{rotating,float,subfigure,graphics,threeparttable}
\usepackage{natbib}

\journal{International Journal of Heat and Mass Transfer}

\newcommand\lae[1]{\label{#1}}

\begin{document}
\begin{frontmatter}
\title{Radiometric force on a sphere in a rarefied gas flow based on the
Cercignani-Lampis model of gas-surface interaction.}

\author[myaddress]{Denize Kalempa}

\author[sharaddress]{Felix Sharipov}

\address[myaddress]{Departamento de Ci\^encias B\'asicas e Ambientais, Escola de
Engenharia de Lorena, Universidade de S\~ao Paulo, 12602-810, Lorena, Brazil}

\address[sharaddress]{Departamento de F\'{\i}sica, Universidade Federal do
Paran\'a, Caixa Postal 19044, 81531-990, Curitiba, Brazil}

\begin{abstract}

The radiometric force on a sphere due to its thermal
polarization in a rarefied gas flow being in equilibrium is investigated on
the basis of a kinetic model to the
linearized Boltzmann equation. The scattering kernel proposed by Cercignani and Lampis
to model the gas-surface interaction using two accommodation coefficients,
namely the tangential momentum accommodation coefficient and the normal
energy accommodation coefficient, is employed as the boundary condition. The
radiometric force on the sphere, as well as the flow field of the gas around
it, are calculated in a wide range of the gas rarefaction,
 defined as the ratio of the sphere radius to an equivalent free path of gaseous particles, 
 covering the free molecular, transition and continuum regimes. The discrete velocity method is employed to solve the kinetic equation
numerically. The calculations are carried out for values of accommodation
coefficients considering most situations encountered in practice. To confirm 
the reliability of the calculations, the reciprocity relation between the cross
phenomena is verified numerically within a numerical error of 0.1\%. 
 The temperature drop between two diametrically opposite points of the spherical
surface in the direction of the gas flow stream, which characterizes the 
thermal polarization effect, is compared to
experimental data for a spherical particle of Pyrex glass immersed in helium
and argon gases. 

\end{abstract}


\end{frontmatter}

\clearpage
\section{Introduction}

In the framework of rarefied gas dynamics, it is well known that a steady gas flow can be induced by 
a temperature field in the absence of external forces such as gravity, e.g. the thermal creep flow in the 
vicinity of a boundary with a longitudinal temperature distribution
\cite{Ken01,Son02,Ann01,Loy07}. Nonetheless, the gas motion can also induce a
thermal effect, e.g. the thermal polarization of a body in a uniform gas
flow \cite{Bak02,Bak05} which leads to the so called radiometric force on
the body. The nonequilibrium phenomena arising within the Knudsen layer are peculiar to rarefied gases and cannot be predicted by the classical equations of 
fluid mechanics \cite{Lan05}. The Knudsen number, $Kn$, defined as
the ratio of the molecular mean free path to a characteristic length scale
of the gas flow, is the parameter used to classify the gas flow
regime as free molecular ($Kn \gg 1$), transitional $(Kn \sim 1)$ and continuum
($Kn \ll 1$). For instance, in problems concerning the transport of small particles in a rarefied gas,
such as air at the standard conditions where the molecular mean free path is
approximately 0.065 $\mu$m, the Knudsen number varies from approximately
zero to 65 when the size of particles ranges from 100 $\mu$m to 10$^{-3}$ $\mu$m. Therefore, the equations of continuum mechanics are not
 valid to describe the gas flow around aerosols in the atmosphere as well as in
many applications which rely on aerosols dynamics.
Moreover, even in the continuum regime there are phenomena which cannot be predicted by the
Navier-Stokes-Fourier equations, e.g. the negative thermophoresis 
in aerosols with a high thermal conductivity related to that of the carrier gas.
The negative thermophoresis was first predicted theoretically by Sone
\cite{Son04} as a result of the thermal stress
slip flow, which is an effect of second order in the Knudsen number.
Experimentally, the negative thermophoresis was already
detected, see e.g. Ref. \cite{Bos02}, in which the
thermophoretic force on a copper sphere in argon gas was measured in a wide
range of the Knudsen number. The radiometric force which appears in a body
placed in an equilibrium gas is also an effect of second
order in the Knudsen number and it is not predicted by the classical equations of
continuum mechanics. Although the magnitude of the radiometric force on a
particle in a uniform gas flow is negligible compared to the viscous drag
force, this force plays an important role in thermophoresis of particles
with arbitrary thermal conductivity. Thus, problems concerning the movement of small
particles in rarefied gases must be solved at the microscopic level by employing the methods of 
rarefied gas dynamics, which are based on the solution of the Boltzmann equation
\cite{CerB2,Sha02B} and 
its related kinetic models, such as the models proposed by Bhatnagar, Gross and Krook \cite{Bha01}
and Shakhov \cite{Shk02}, or on the direct simulation Monte Carlo method \cite{Bir02}.

Historically, the term radiometric force is widely known from the 
experiments designed by Crookes in the 19th century concerning the rotation of a windmill 
in a closed vessel by the incidence of light \cite{Crookes01}. Since the Crooke's theory of pressure 
radiation was not sucessfull to provide an overall explanation for the physics underlying his experiments, 
other theories were proposed over the years by prominent scientists such as Reynolds, Maxwell and Einstein, 
see e.g. Refs. \cite{Crookes02,Crookes03}. However, the more reasonable explanation 
for the origin of the force in the Crookes radiometer were those proposed by
Maxwell \cite{Crookes04} and Reynolds \cite{Crookes05} which rely on kinetic theory of
gases \cite{Ken01}. A more recent historical review on radiometric
phenomena, including some applications, is given in Ref. \cite{Ket02}.
It is worth mentioning that the increasing interest in numerical and
experimental studies concerning the radiometric force is
due to the promising applications in the field of micro and nano technology,
see e.g. the literature review on Knudsen pumps \cite{2020Knpumps} and Refs.
\cite{Gerd,Ger1} concerning gas sensors. The description of the main mechanisms
of thermally induced flows which lead to the radiometric force
in rarefied gas systems is given in Ref. \cite{Son20}. It is important to point out that there are many
possibilities to build a configuration in which the radiometric force
arises. In thin plates it is usually called Knudsen force due to the
pioneering works by Knudsen, e.g. Ref. \cite{Knu02}. Small particles in a rarefied gas 
illuminated by a beam of light or other
radiation also experience a radiometric force as well as a radiation
pressure force. In this case, the non-uniform heating of particles is caused by the absorption 
of electromagnetic energy and the force is called as photophoretic. Thus, studies concerning 
radiometric phenomena in rarefied gases are also important to understand the physics underlying the transport 
of aerosols in the atmosphere due to absorption of solar radiation and in applications such as the optical
trapping and manipulation of small particles \cite{Zemanek:19}. According to Ref. \cite{Zemanek:19}, 
in a gas medium the photophoretic force dominates the optical manipulation of light-absorbing 
particles because its order of magnitude is larger than that corresponding to the
pressure-radiation.
 
 Usually, the classical problem of viscous drag on a sphere relies on the assumption 
of uniform temperature on the spherical surface, which is valid in case of a sphere with high thermal
conductivity related to that of the carrier gas. Otherwise, for low and
moderate thermal conductivity of the solid particle, the non-uniform temperature of
its surface due to the thermal polarization in a uniform gas flow must be considered in the boundary
condition. The temperature non-uniformity of the spherical particle leads to nonequilibrium phenomena 
in the vicinity of the boundary from which arise the radiometric force on the
sphere. This force as well as the macroscopic characteristics of the gas
flow around the sphere induced by
the temperature field are strongly dependent on the
gas-surface interaction law and accommodation coefficients at the spherical
surface. In fact, experiments on thermal
polarization of small particles can be carried out to determine the values of the 
accommodation coefficients, see e.g. Ref. \cite{Bak05}. Thus, the proper modelling of
the gas flow around the sphere and the calculation of the force acting on it depends on
the mathematical model of gas-surface interaction and correct
values of accommodation coefficients. Currently, there are few papers in the 
literature concerning the influence of the gas-surface interaction law on the radiometric
phenomenon in the whole range of the gas rarefaction and most of them are restricted to the
 assumption of diffuse scattering or complete accommodation of gas molecules on the surface.
To the best of our knowledge, Beresnev and coworkers \cite{Ber10} were the
first authors to analyse the influence of the accommodation coefficients on
the radiometric force on a sphere in a uniform gas flow by solving a kinetic
model to the Boltzmann equation in the whole range of the Knudsen
number. As pointed out in our previous work \cite{Sha130}, in the boundary
condition used in Ref. \cite{Ber10}, the distribution function of reflected
molecules was expanded in Hermite polynomials and the unknown accommodation
coefficients of momentum and energy were determined from the conservation
laws of momentum and energy on the surface. Moreover, the variational method
was applied by the authors to solve the kinetic equation. The results presented in Ref.
\cite{Ber10} for the temperature difference between the two diametrically opposite
points of the spherical surface in the direction of the gas flow stream,
which characterizes the thermal polarization, showed a strong dependence on
the accommodation coefficients. 

An interesting feature of rarefied gas flows induced by temperature fields
is the absence of gas motion in the free molecular regime when the
diffuse-specular model of gas-surface interaction proposed by Maxwell
\cite{Max01} is used in the boundary condition, see e.g. Refs. \cite{1984Sone,Cer44,Tak08}.
The model proposed by Maxwell assumes that only a part of the gas molecules
is reflected diffusely, while the remaining part is reflected specularly. In spite of widely
used due to its mathematical simplicity, the Maxwell model has some drawnbacks. For instance,
the Maxwell model cannot predict the correct exponent which appears in the
thermomolecular pressure difference (TPD) in the free molecular regime. For
instance, while many experiments lead to an TPD exponent varying from 0.4 to 0.5,  see
e.g. Refs. \cite{Pod01,Edm01}, the Maxwell model always provides a value of 0.5. 
To verify if a steady gas flow is induced by a temperature field in
the free molecular regime when other model of gas-surface interaction is
used, Kosuge \textit{et al.} \cite{Kos09} carried out the calculations for the problem of a
rarefied gas between parallel plates with non-uniform temperature by
employing the model of gas-surface interaction proposed by Cercignani and Lampis
(CL) \cite{Cer11}. In spite of the mathematical complexity, the CL model of gas-surface interaction 
allows the setting of two accommodation
coefficients, namely the tangential momentum accommodation coefficient
(TMAC) and the normal energy accommodation coefficient (NEAC). Currently, the values of
the NEAC and TMAC extracted from experiments can be found in the literature for
several gases and surfaces, see e.g. \cite{Sem01,Sha113,Tro01,Sha36}. For
instance, the NEAC ranges from 0 to 0.1 for helium and from 0.5 to 0.95 for
argon at ambient temperature and metallic surfaces such as aluminum,
platinum and stainless steel, while the TMAC ranges from 0.5 to 0.95 for
both gases at the same conditions. According to
the results presented in Ref. \cite{Kos09}, obtained by applying a deterministic
method to solve the integral equation 
derived from the Boltzmann equation in the free molecular limit, a steady gas flow is induced by the 
temperature field in case of non-difuse scattering of gas molecules on the surface.
In fact, different gas flow patterns were observed between the plates by varying the accommodation
coefficients. 

 In our previous paper \cite{Sha130}, the CL model of gas-surface
interaction was employed in the modelling of the viscous drag and thermophoresis
on a sphere in a rarefied gas. The case of particle with high thermal 
conductivity related to that of the carrier gas
was considered so that the temperature of the sphere was assumed as being a
constant. The modelling was based on the numerical solution of the Shakhov
model \cite{Shk02} for the linearized Boltzmann equation via the discrete
velocity method. The drag and thermophoretic forces on the sphere, as well as the flow
fields around it, were obtained in a wide range of the Knudsen number and accommodation
coefficients. According to the previous results, both forces are sensitive to the
accommodation coefficients. Indeed, for some sets of accommodation coefficients,
the appearance of the negative thermophoresis was observed in the near continuum
regime. 

In the present work, the problem of radiometric force on a spherical particle
caused by its thermal polarization in a uniform gas flow is investigated
numerically with basis on the kinetic model proposed by Shakhov \cite{Shk02} for the
linearized Boltzmann equation and the discrete velocity method. As already
mentioned, the magnitude of this force is negligible compared to the viscous
drag force, but it plays an important role in thermophoresis of particles
immersed in a rarefied gas. It is worth mentioning that nowadays, in spite
of the great computational infrastructure available, the kinetic models are still
widely used because their solutions require a modest computational effort
compared to that required to solve the exact Boltzmann equation. The kinetic model
proposed by Shakhov is considered the most reliable to deal with problems concerning both heat and mass transfer
in a single gas because it provides the correct Prandtl number, i.e. the correct values for
the viscosity and heat conductivity of the gas, and maintains the original
properties of the Boltzmann equation corresponding to mass, momentum and
energy conservation laws, and the H-theorem. Moreover, it provides a good accuracy
and its reliability is supported by the literature. For instance, for the
classical problems of Couette flow and heat transfer between parallel
plates, the comparison between the results obtained from the Shakhov model
and those obtained from the Boltzmann equation for hard-spheres potential is
presented in Refs. \cite{Sha02B, Gra06,Sha131} and shows that the difference is
within 5\%. In the context of forces on small particles in a rarefied gas, a
similar comparison is presented in Refs. \cite{Ber09,Ber10,Sha130} and shows that
the difference is within 5-7\%. The calculations are carried out in a range of the Knudsen number
 which covers the free molecular, transition and continuum regimes.
Moreover, a wide range of the NEAC and TMAC is considered. The fullfillment
of the reciprocity relation between cross phenomena is verified. As results,
the radiometric force on the sphere and the flow fields around it induced by
the temperature field are presented. The temperature drop between the two diametrically opposite points
in the sphere in the direction of the gas flow stream is calculated and the
results are compared to the experimental data provided in Ref. \cite{Bak05} 
for a Pyrex glass sphere immersed in helium and argon gas. 

\section{Statement of the problem}

\begin{figure}[ht]
\centering
\includegraphics[scale=1]{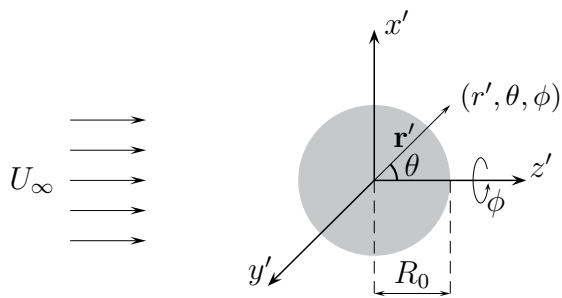}
\caption{Scheme of the problem}
\label{fig1}
\end{figure}

Let us consider a sphere of radius $R_0$ and thermal conductivity $\lambda_p$ 
at rest placed in a monoatomic rarefied gas of thermal conductivity
$\lambda_g$. Far from the sphere, the gas
flows with a constant bulk velocity $U_{\infty}$ in the $z'$-direction as
showed in Figure \ref{fig1}. Moreover, far from the sphere the equilibrium 
 gas number density, temperature and pressure are denoted by $n_0$, $T_0$ and
$p_0$, respectively, and these quantities are related by the state equation
$p_0$=$n_0kT_0$, where $k$ denotes the Boltzmann constant. 

Due to the geometry of the problem, spherical coordinates $(r',\theta,\phi)$ in the
physical space are introduced so that the components of the position
vector ${\bf r}'$ of gaseous particles read
\begin{subequations}
\begin{align}
&x'=r'\sin{\theta}\cos{\phi},\\
&y'=r'\sin{\theta}\sin{\phi},\\
&z'=r'\cos{\theta}.
\end{align}
\lae{sp1}
\end{subequations}
Moreover, the components of the molecular velocity vector ${\bf v}$ are
given as
\begin{subequations}
\begin{align}
&v_x=(v_r\sin{\theta}+v_{\theta}\cos{\theta})\cos{\phi}-v_{\phi}\sin{\phi},\\
&v_y=(v_r\sin{\theta}+v_{\theta}\cos{\theta})\sin{\phi}+v_{\phi}\cos{\phi},\\
&v_z=v_r\cos{\theta}-v_{\theta}\sin{\theta},
\end{align}
\lae{sp2}
\end{subequations} 
where $v_r$, $v_{\theta}$ and $v_{\phi}$ are the radial, polar and azimuthal
components of the molecular velocity vector, respectively, which are written
in spherical coordinates $(v,\theta',\phi')$ in the velocity space as follows
\begin{subequations}
\begin{align}
&v_r=v\cos{\theta'},\\
&v_{\theta}=v_t\cos{\phi'},\\
&v_{\phi}=v_t\sin{\phi'},
\end{align}
\lae{sp3}
\end{subequations}
with the tangential component given as
\begin{equation}
v_t=\sqrt{v_{\theta}^2+v_{\phi}^2}=v\sin{\theta'}.
\lae{sp4}
\end{equation}
For convenience, hereafter, the dimensionless sphere
radius, $r_0$, as well as the position ${\bf r}$ and molecular velocity ${\bf c}$
vectors are introduced as
\begin{equation}
r_0=\frac{R_0}{\ell_0},\quad {\bf r}=\frac{{\bf r}'}{\ell_0},\quad {\bf
c}=\frac{{\bf v}}{v_0},
\lae{sp5}
\end{equation}
where $\ell_0$ and $v_0$ denote the equivalent free path and the most
probable molecular speed, defined as
\begin{equation}
\ell_0=\frac{\mu_0v_0}{p_0}, \quad
\quad v_0=\sqrt{\frac{2kT_0}{m}}.
\lae{sp6}
\end{equation}
Here, $\mu_0$ denotes the viscosity of the gas at temperature
$T_0$, while $m$ is the molecular mass.

It is assumed that the ratio of thermal conductivities,
$\Lambda$=$\lambda_p/\lambda_g$, is arbitrary. As a consequence, the
non-uniform temperature of the spherical surface must be taken into account in the solution of
the problem because it induces a gas flow around 
the sphere as well as a radiometric force on the sphere. 
Thus, the temperature of the spherical particle, $T_p$, is obtained from the solution of the Laplace equation 
for heat conduction with axial simmetry written as 
\begin{equation}
\biggl[
\frac{\partial^2}{\partial r^2}+\frac 2r \frac{\partial }{\partial r}
+\frac{1}{r^2\sin{\theta}}\frac{\partial}{\partial \theta}\left(\sin{\theta}
\frac{\partial}{\partial \theta}\right)\biggr]
T_p(r,\theta)=0.
\label{la1}
\end{equation}
The solution of equation (\ref{la1}) must satisfy two conditions. Firstly, the continuity of the
radial component of the heat flux at the gas-solid interface which reads
\begin{equation}
q_r(r_0,\theta)=-\frac{15}{8}\frac{\Lambda}{T_0} \frac{\partial T_p}{\partial
r},
\lae{la2}
\end{equation}
where the dimensionless radial component of the heat flux was introduced as
\begin{equation}
q_r=\frac{Q_r}{p_0v_0},
\lae{la3}
\end{equation}
with $Q_r$ denoting the corresponding dimensional quantity. Secondly, the condition of 
finite temperature at the center of the spherical particle, i.e. at $r$=0, must be
satisfied. Thus, the solution of the heat conduction equation (\ref{la1})
satisfying both conditions leads to the following temperature distribution at the
spherical surface of the solid particle 
\begin{equation}
T_p(r_0,\theta)=T_0(1+\tau_{s0}\cos{\theta}),
\lae{la4}
\end{equation} 
where $\tau_{s0}$ is a constant which depends on the gas flow stream and
heat conductivity of the particle. Note that the same temperature distribution on the particle
surface can be caused by a light radiation. In this case, $\tau_{s0}$
depends on the light intensity and heat conductivity of the particle. Thus,
the solution presented below is determined by the distribution (\ref{la4}),
but it is independent of the physical phenomenon leading to the
distribution. 
 
Two dimensionless thermodynamic forces are introduced as 
\begin{equation}
X_u=\frac{U_{\infty}}{v_0} ,\quad
X_q=\tau_{s0}.
\lae{sp9}
\end{equation}
It is assumed a weak disturbance from thermodynamic 
equilibrium, which means 
\begin{equation}
|X_u|\ll 1, \quad |X_q|\ll 1.
\lae{sp9a}
\end{equation}

The thermodynamic forces are coupled due to the effect of thermal
polarization of the particle in the moving gas. However, the assumptions given in (\ref{sp9a}) 
allow us to decompose the problem 
into two independent parts corresponding to viscous drag on a sphere
with uniform temperature and radiometric force on a sphere with non-uniform
temperature placed in an equilibrium gas. Thus, after solving the problem
separately for each thermodynamic force, a solution describing the real
situation is given as a superposition of the obtained solutions. Note 
that the radial heat flux $q_r(r_0,\theta)$, given in (\ref{la2}),
accounts to the isothermal heat transfer due to the gas motion as well as
 to the non-isothermal heat transfer due to the temperature difference between
the particle and the gas around it.  


The main parameter determining the solution of the problem is the rarefaction parameter, $\delta$,
which is inversely proportional to the Knudsen number, but defined
here as the ratio of the sphere radius to the equivalent molecular free path, i.e.
\begin{equation}
\delta=\frac{R_0}{\ell_0},
\lae{a7}
\end{equation}
where $\ell_0$ is defined in (\ref{sp6}). Note that, according to
(\ref{sp5}), $r_0$=$\delta$. When $\delta \ll 1$ the gas is in the free molecular regime, while the
opposite limit, $\delta \gg 1$, corresponds to the continuum or hydrodynamic
regime. In other situations, $\delta \sim 1$, the gas is in the transition regime.

The CL-model of the gas-surface interaction \cite{Cer11} is employed in the boundary
 condition. In this model, the type of the
gas-surface interaction is chosen by setting appropriate values for the NEAC
and TMAC. Henceforth, these accommodation coefficients will be denoted by
$\alpha_n$ and $\alpha_t$, respectively. The widely used diffuse scattering or complete
accommodation on the surface corresponds to $\alpha_n$=1 and $\alpha_t$=1.  

The influence of the NEAC and TMAC on the
viscous drag on a sphere with high thermal
conductivity related to that of the carrier gas, which corresponds to the
case of a sphere with uniform temperature, was already investigated in
our previous work \cite{Sha130} in a range of the gas 
rarefaction which covers the free molecular, 
transition and hydrodynamic regimes. Thus, in the present work, focus is given
to the solution due to the thermodynamic force $X_q$. We are going to
calculate the radiometric force on
the sphere and the flow fields around it induced by $X_q$ in a wide range of the gas
rarefaction and accommodation coefficients. It is worth mentioning that the radiometric force is
negligible compared to the viscous drag force on the sphere, but the introduction of two
thermodynamic forces as defined in (\ref{sp9}) allows us to verify the fullfillment of the reciprocity relation
between cross phenomena as an additional criterium for the accuracy of the
numerical calculations.

\section{Kinetic equation}

Similarly to our previous work \cite{Sha130}, the model proposed by Shakhov \cite{Shk02}
for the Boltzmann equation is employed here due to its reliability to deal with
problems concerning both mass and heat transfer. Thus, the kinetic equation
is written as
\begin{equation}
{\bf v}\cdot \frac{\partial f}{\partial {\bf r}'}=Q(ff_*),
\lae{ke0}
\end{equation}
where $f$=$f({\bf r}',{\bf v})$ is the distribution function of molecular
velocities and
\begin{equation}
Q(ff_*)=\nu_S\left\{f^M\left[1+\frac{4}{15}\left(\frac{V^2}{v_0^2}-\frac 52\right)
\frac{{\bf Q}\cdot {\bf V}}{p_0v_0^2}\right]-f({\bf r}',{\bf v}) \right\}
\lae{ke4}
\end{equation}
is the intermolecular collision integral.The local Maxwellian function reads
\begin{equation}
f^M({\bf r}',{\bf v})=n\left[\frac{m}{2\pi kT({\bf
r}')}\right]^{3/2}\exp{\left[-\frac{m{\bf V}^2}{2kT({\bf r}')}\right]},
\lae{ke4a}
\end{equation}
the quantity $\nu_S$ has the order of the
intermolecular interaction frequency and ${\bf V}$=${\bf v}-{\bf U}$ is the
peculiar velocity so that $V=|{\bf V}|$ denotes its magnitude. ${\bf U}({\bf r}')$ and ${\bf
Q}({\bf r}')$ are the bulk velocity and heat flux vectors. 

The assumptions of smallness of the thermodynamic forces $X_u$ and $X_q$, defined in
(\ref{sp9}), allow us to linearize the kinetic equation by representing the
distribution function of molecular velocities as
\begin{equation}
f({\bf r},{\bf c})=f_R^{M}[1+h^{(u)}({\bf r},{\bf v})X_u +
h^{(q)}({\bf r},{\bf v})X_q],
\lae{ke8}
\end{equation}
where $h^{(u)}$ and $h^{(q)}$ are the perturbation functions due to the
thermodynamic forces $X_u$ and $X_q$, respectively. The reference Maxwellian function is
given by the equilibrium distribution function far from the sphere, i.e.
\begin{equation}
f_R^{M}=f_{\infty}^{M}=f_0\left(1+2c_zX_u\right),
\lae{ke9}
\end{equation}
where $f_0$ is the global Maxwellian function.

 The representation (\ref{ke8}) allows to write the gas number density,
the temperature, the dimensionless bulk velocity and heat flux vectors as
follows 
\begin{equation}
n(r,\theta)=n_0[1+\nu^{(u)}(r,\theta)X_u+\nu^{(q)}(r,\theta)X_q],
\lae{ke12a1}
\end{equation}
\begin{equation}
T(r,\theta)=T_0[1+\tau^{(u)}(r,\theta)X_u+\tau^{(q)}(r,\theta)X_q],
\lae{ke12b1}
\end{equation}
\begin{equation}
{\bf u}(r,\theta)=\frac{{\bf U}}{v_0}={\bf u}^{(u)}(r,\theta)X_u + {\bf
u}^{(q)}(r,\theta)X_q,
\lae{ke12c1}
\end{equation}
\begin{equation}
 {\bf q}(r,\theta)=\frac{{\bf Q}}{p_0v_0}={\bf q}^{(u)}(r,\theta)X_u + {\bf
q}^{(q)}(r,\theta)X_q,
\lae{ke12d1}
\end{equation}
where $\nu^{(n)}$ and $\tau^{(n)}$ are the density and
temperature deviations from equilibrium, while ${\bf u}^{(n)}$ and ${\bf
q}^{(n)}$ are the bulk velocity and heat
flux due to the corresponding thermodynamic force ($n$=$u, q$).

Thus, after substituting the representation (\ref{ke8}), the
dimensionless quantities defined in (\ref{sp5}) and the macroscopic
quantities (\ref{ke12a1})-(\ref{ke12d1}) into (\ref{ke0}), 
the linearized kinetic equation for each thermodynamic force is written as
\begin{equation}
\hat{D}h^{(n)}=\hat{L}_S h^{(n)},\quad n=u, q.
\lae{ke10}
\end{equation}
Due to the spherical geometry of the problem, the kinetic equation (\ref{ke10})
is written in spherical coordinates ($r,\theta,\phi$) in the physical space as well as in
the molecular velocity space ($c,\theta',\phi'$). Details regarding this transformation 
can be found in Ref. \cite{Shk14}. Thus, after taking into account the simmetry of
the problem on the azimuthal angle $\phi$, the transport operator $\hat{D}$
and the linearized collision integral $\hat{L}_S$ appearing in the kinetic
equation (\ref{ke10}) are written as 
\begin{equation}
\hat{D}h^{(n)}=
c_r\frac{\partial h^{(n)}}{\partial r}- \frac{c_t}{r}\frac{\partial
h^{(n)}}{\partial \theta'}+\frac{c_t}{r}\cos{\phi'}\frac{\partial
h^{(n)}}{\partial \theta}-
\frac{c_t}{r}\sin{\phi'}\cot{\theta}\frac{\partial
h^{(n)}}{\partial \phi'}
\lae{ke11}
\end{equation}
and 
\begin{equation}
\hat{L}_Sh^{(n)}
=\nu^{(n)}+\left(c^2-\frac
32\right)\tau^{(n)}+2{\bf c}\cdot{\bf u}^{(n)}+\frac{4}{15}\left(c^2-\frac
52\right){\bf c}\cdot {\bf q}^{(n)}-h^{(n)},
\lae{ke12}
\end{equation}
where $h^{(n)}$=$h^{(n)}(r,\theta,{\bf c})$. 

The macroscopic characteristics (\ref{ke12a1})-(\ref{ke12d1}) are calculated as moments
of the distribution function of molecular velocities and details concerning
such a calculation can be found in Ref. \cite{Fer02}. In our notation, the
density and temperature deviations from equilibrium read
\begin{equation}
\nu^{(n)}(r,\theta)=\frac{1}{\pi^{3/2}}\int h^{(n)}(r,\theta,{\bf
c})\mbox{e}^{-c^2}\, d{\bf}{\bf c},
\lae{k12a}
\end{equation}
\begin{equation}
\tau^{(n)}(r,\theta)=\frac{2}{3\pi^{3/2}}\int \left(c^2 -\frac
32\right)h^{(n)}(r,\theta,{\bf
c})\mbox{e}^{-c^2}\, d{\bf}{\bf c},
\lae{k12b}
\end{equation}
while the radial an polar components of the bulk velocity and heat flux
vectors are given as 
\begin{equation}
u_r^{(n)}(r,\theta)=\frac{1}{\pi^{3/2}}\int c_rh^{(n)}(r,\theta,{\bf
c})\mbox{e}^{-c^2}\, d{\bf}{\bf c},
\lae{k12d}
\end{equation}
\begin{equation}
u_{\theta}^{(n)}(r,\theta)=
\frac{1}{\pi^{3/2}}\int c_{\theta}h^{(n)}(r,\theta,{\bf
c})\mbox{e}^{-c^2}\, d{\bf}{\bf c},
\lae{k12e}
\end{equation}
\begin{equation}
q_r^{(n)}(r,\theta)=\frac{1}{\pi^{3/2}}\int c_r\left(c^2-\frac 52\right)h^{(n)}(r,\theta,{\bf
c})\mbox{e}^{-c^2}\, d{\bf}{\bf c},
\lae{k12f}
\end{equation}
\begin{equation}
q_{\theta}^{(n)}(r,\theta)=\frac{1}{\pi^{3/2}}\int c_{\theta}\left(c^2-\frac 52\right)h^{(n)}(r,\theta,{\bf
c})\mbox{e}^{-c^2}\, d{\bf}{\bf c},
\lae{k12g}
\end{equation}
where $d{\bf c}$=$c^2\sin{\theta'}\mbox{d}c\mbox{d}\theta'\mbox{d}\phi'$. The force acting on 
the sphere in the $z'$-direction is calculated from the normal and
tangential stress on the sphere, whose expressions can be found in Ref.
\cite{Fer02}. Here, the dimensionless force in the $z$-direction is
introduced as 
\begin{equation}
F_z=\frac{F_z'}{4\pi R_0^2p_0}=F_uX_u + F_qX_q,
\lae{ke13}
\end{equation}  
where the viscous drag force $F_u$ and the radiometric force $F_q$ are given as
\begin{equation}
F_u=-\frac{1}{2\pi^{5/2}}\int_{\Sigma_w}\mbox{d}\Sigma_w
\int c_rc_z\mbox{e}^{-c^2}[h^{(u)}(r_0,\theta,{\bf c})+2c_z]\, \mbox{d}{\bf
c},
\lae{ke14}
\end{equation}
\begin{equation}
F_q=
-\frac{1}{2\pi^{5/2}}\int_{\Sigma_w}\mbox{d}\Sigma_w\int c_rc_z\mbox{e}^{-c^2}
h^{(q)}(r_0,\theta, {\bf c})\, \mbox{d}{\bf c}.
\lae{ke15}
\end{equation}
$\mbox{d}\Sigma_w$=$\sin{\theta}\mbox{d}\theta\mbox{d}\phi$ denotes a
dimensionless area element in the spherical surface. 

Far from the sphere ($r\rightarrow \infty$), the asymptotic behavior of the
perturbation functions are obtained from the Chapman-Enskog solution for the
linearized kinetic equation as
\begin{equation}
h_{\infty}^{(u)}=\lim_{r\rightarrow \infty} h^{(u)}(r, \theta,{\bf c})=0,
\lae{k15a}
\end{equation}
\begin{equation}
h_{\infty}^{(q)}=\lim_{r\rightarrow \infty} h^{(q)}(r, \theta,{\bf c})=0.
\lae{k15b}
\end{equation}

\section{Boundary condition}

The boundary condition to solve the kinetic equation for each thermodynamic
force is obtained from the relation
between the distribution functions of incident particles on the sphere and
reflected particles from the sphere. According to Refs. \cite{CerB2,Sha02B}, 
the general form of the linearized boundary condition at the spherical
surface reads
\begin{equation}
h^{+(n)}=\hat{A}h^{-(n)}+h_w^{(n)}-\hat{A}h_w^{(n)},
\lae{bc1}
\end{equation}
where the signal $``+"$ denotes the reflected particles from the surface, while
the signal $``-"$ denotes the incident particles on the surface. For the
problem in question, the source terms are given as 
\begin{equation}
h_w^{(u)}=-2c_z,\quad 
h_w^{(q)}=\left(c^2-\frac 32\right)\frac{z_0}{\delta}\cos{\theta},
\lae{bc2}
\end{equation}
where $c_z$=$c_r\cos{\theta}-c_{\theta}\sin{\theta}$.

In spherical coordinates, the scattering operator $\hat{A}$ is decomposed as
\begin{equation}
\hat{A}h^{(n)}=\hat{A}_r\hat{A}_{\theta}\hat{A}_{\phi}h^{(n)},
\lae{bc2a}
\end{equation}
where
\begin{equation}
\hat{A}_r\xi=
-\frac{1}{c_r}\int_{c_r'<0}c_r'\exp{(c_r^2-c_r'^{2})}R_r(
c_r\rightarrow c_r')\xi(c_r') \,\mbox{d}c_r',
\lae{bc3}
\end{equation}
\begin{equation}
\hat{A}_i\xi=\int_{-\infty}^{\infty}\exp{(c_i^2-c_i'^2)}R_i(c_i\rightarrow
c_i')\xi(c_i') \, \mbox{d}c_i',\quad i=\theta, \phi,
\lae{bc3a}
\end{equation}
for an arbitrary $\xi$ as function of the molecular velocity. According to the
scattering kernel proposed by Cercignani and Lampis \cite{Cer11},
the functions $R_r$, $R_{\theta}$ and $R_{\phi}$ are given by
\begin{equation}
R_r(c_r\rightarrow
c_r')=\frac{2c_r}{\alpha_n}\exp{\left[-\frac{c_r^2+(1-\alpha_n)c_r'^2}{\alpha_n}\right]}
I_0\left(\frac{2\sqrt{1-\alpha_n}}{\alpha_n}c_rc_r'\right),
\lae{bc5}
\end{equation}
\begin{equation}
R_i(c_i\rightarrow c_i')=\frac{1}{\sqrt{\pi
\alpha_t(2-\alpha_t)}}\exp{\left\{-\frac{[c_i-(1-\alpha_t)
c_i']^2}{\alpha_t(2-\alpha_t)}\right\}},\quad i=\theta, \phi,
\lae{bc6}
\end{equation}
where $I_0$ denotes the modified Bessel function of first kind and zeroth
order. In this model of gas-surface interaction, the accommodation coefficients
 can vary in the ranges $0 \le \alpha_t \le 2$ and $0\le \alpha_n \le 1$. The case 
$\alpha_t$=1 and $\alpha_n$=1 corresponds to diffuse scattering or complete
accommodation on the spherical surface, while the case $\alpha_t$=0 and $\alpha_n$=0
corresponds to specular reflection at the surface.
 
After some algebraic manipulation, it can be shown that
\begin{equation}
\hat{A}_ic_i=(1-\alpha_t)c_i,\quad i=\theta, \phi,
\lae{bc7}
\end{equation}
\begin{equation}
\hat{A}_ic_i^2=(1-\alpha_t)^2c_i^2+\frac 12 \alpha_t(2-\alpha_t),
\lae{bc8}
\end{equation}
\begin{equation}
\hat{A}_rc_r=-\sqrt{\alpha_n}H_1(\eta),
\lae{bc10}
\end{equation}
\begin{equation}
\hat{A}_rc_r^2=\alpha_n + (1-\alpha_n)c_r^2,
\lae{bc11}
\end{equation}
where
\begin{equation}
H_1(\eta)=2\mbox{e}^{-\eta^2}\int_{0}^{\infty}\xi^{2}\mbox{e}^{-\xi^2}I_0(2\eta
\xi)\, \mbox{d}\xi,\quad \xi=\frac{c_r'}{\sqrt{\alpha_n}},
\lae{bc13}
\end{equation}
and 
\begin{equation}
\eta=c_r\sqrt{\frac{1}{\alpha_n}-1}.
\lae{bc12a}
\end{equation}

Therefore, from (\ref{bc1}), the boundary conditions at $r$=$r_0$ for the perturbation
functions of reflected gas particles, i.e. $c_r >0$, from the spherical surface are written
as
\begin{equation}
h^{+(u)}=\hat{A}h^{-(u)}-2\frac{z_0}{\delta}[(1-\alpha_t)
c_r+\sqrt{\alpha_n}H_1(\eta)]-2\alpha_tc_z,
\lae{bc12b}
\end{equation}
\begin{equation}
h^{+(q)}=\hat{A}h^{-(q)}-\frac{z_0}{\delta}[\alpha_n(1-c_r^2)+\alpha_t(2-\alpha_t)(1-c_t^2)].
\lae{bc12c}
\end{equation}

\section{Reciprocity relation}

In our previous paper \cite{Sha130}, the reciprocal relation between the thermophoretic force 
and drag force solution was obtained in its explicit form. The analogous relation between the 
radiometric force and drag force solution is simpler and follows directly from the theory described 
in Ref.\cite{Sha12}. Then, the explicit relation was obtained in the subsequent work, 
see Eq.(5.42) from \cite{Sha13}, which in our dimensionless notations reads
\begin{equation}
4\pi  F_q=-\int_{\Sigma_w} \tau_p  q_r^{(u)}(r_0,\theta) \mbox{d} \Sigma_w
\label{AA}
\end{equation}
where the temperature deviation of the particle surface $\tau_p$ 
 is calculated from (\ref{la4}) and (\ref{sp9}) as
\begin{equation}
\tau_p=\frac{T_p-T_0}{X_q T_0}=\cos\theta. 
\end{equation}
Considering that $\mbox{d} \Sigma_w= \sin\theta \mbox{d}\theta \mbox{d}\phi$ and integrating (\ref{AA}) 
with respect to $\phi$, the reciprocal relation is reduced to  
\begin{equation}
F_q=-\frac 12\int_0^\pi q_r^{(u)}(r_0,\theta) \cos\theta \sin\theta \mbox{d}\theta.
\lae{AC}
\end{equation}
According to Eq.(6.16) from \cite{Sha130}, $q_r^{(u)}$ can be represented as
\begin{equation}
q_r^{(u)}(r,\theta)=q_r^{*(u)}(r)\cos\theta.
\lae{AB}
\end{equation}
Substituting (\ref{AB}) into (\ref{AC}), we obtain the reciprocal relation in its simple form 
\begin{equation}
F_q=-\frac 13 q_r^{*(u)}(r_0).
\lae{AB1}
\end{equation}

\section{Force in the free molecular and continuum regimes}

In the free molecular molecular regime, i.e. $\delta \ll 1$, the
perturbation function of incident gas particles on the surface is not
perturbed, which means that $h^{-(n)}$=$h_{\infty}^{(n)}$ for the
corresponding thermodynamic force. Thus, from (\ref{k15a}) and (\ref{k15b}),
$h^{-(n)}$=0 for both thermodynamic forces. Therefore, the substitution of the solution
(\ref{bc12c}) into (\ref{ke15}) leads to the following expression for the
radiometric force on the sphere in the free molecular regime
\begin{equation}
F_q=-\frac{\alpha_n}{12},
\label{new1}
\end{equation}
which depends only on the NEAC. 

In the continuum regime, i.e. $\delta \gg 1$, the radiometric force is
obtained from the Navier-Stokes equations with thermal slip boundary
condition. Thus, according to Ref. \cite{Sha13}, the radiometric force on
the sphere reads
\begin{equation}
F_q=-\frac{\sigma_T}{2\delta^2},
\label{new2}
\end{equation}
where $\sigma_T$ is the thermal slip coefficient,
 which depends on the NEAC and TMAC as well as on the intermolecular interaction potential. Reliable
values for the thermal slip coefficient can be found in Ref.\cite{Sha84}.

The forces (\ref{new1}) and (\ref{new2}) satisfy the reciprocity relation
(\ref{AB1}). In the free molecular regime, the radial component of the heat
flux at the boundary is obtained just by substituting (\ref{bc12b}) into
(\ref{k12f}). After some algebraic manipulation the following expression is
obtained
\begin{equation}
q_r^{(u)}(r_0)=\frac{\alpha_n}{4}.
\label{new3}
\end{equation}

In the continuum regime, the heat flux cannot be obtained 
by the classical Navier-Stokes-Fourier equations because it 
appears in an approximation of second order in the Knudsen number, see e.g. \cite{Fer02,Cha04}. 
Thus, the reciprocity relation (\ref{AB1}) is verified numerically by
comparing the present results obtained for the radiometric force with those obtained
previously \cite{Sha130} for the heat flux.

\section{Numerical solution}

Like in our previous paper \cite{Sha130}, in order to eliminate the dependence of the 
numerical solution on the variables $\theta$ and $\phi'$, 
the similarity solution proposed in Ref. \cite{Son32} 
is employed. Thus, the dependence of the numerical scheme on the variables
$\theta$ and $\phi'$ is eliminated by representing the perturbation function $h^{(q)}$ as
\begin{equation}
h^{(q)}(r,\theta, {\bf c})=h_c^{(q)}(r,c,\theta')\cos{\theta} +
h_s^{(q)}(r,c,\theta')c_{\theta}\sin{\theta}.
\lae{nm1}
\end{equation}
The substitution of this representation into the kinetic equation (\ref{ke10})
for the thermodynamic force $X_q$ leads to the following system of equations
for the functions $h_c^{(q)}$ and $h_s^{(q)}$
\[
c_r\frac{\partial h_c^{(q)}}{\partial r}-\frac{c_t}{r}\frac{\partial
h_c^{(q)}}{\partial
\theta'}+\frac{c_t^2}{r}h_s^{(q)}=\nu^{*(q)}+\left(c^2-\frac
32\right)\tau^{*(q)}+2c_ru_r^{*(q)}
\]
\begin{equation}
\hskip2cm
+\frac{4}{15}c_r\left(c^2-\frac 52 \right)q_r^{*(q)}-h_c^{(q)},
\lae{nm1a1}
\end{equation}
\[
c_r\frac{\partial h_s^{(q)}}{\partial r}-\frac{c_t}{r}\frac{\partial
h_s^{(q)}}{\partial \theta'}-\frac{c_r}{r}h_s^{(q)}-\frac
1rh_c^{(q)}=2u_{\theta}^{*(q)}
\]
\begin{equation}
\hskip2cm
+\frac{4}{15}\left(c^2-\frac 52
\right)q_{\theta}^{*(q)}-h_s^{(q)},
\lae{nm1a2}
\end{equation}
where the quantities with $*$ are calculated from (\ref{k12a})-(\ref{k12g})
as
\begin{equation}
\nu^{*(q)}(r)=\frac{\nu^{(q)}(r,\theta)}{\cos{\theta}}=
\frac{2}{\sqrt{\pi}}\int_{0}^{\infty}\int_{0}^{\pi}c_th_c^{(q)}
\mbox{e}^{-c^2}\, c\mbox{d}c\mbox{d}\theta',
\lae{apb5}
\end{equation}
\begin{equation}
\tau^{*(q)}(r)=\frac{\tau^{(q)}(r,\theta)}{\cos{\theta}}=
\frac{4}{3\sqrt{\pi}}\int_{0}^{\infty}\int_{0}^{\pi}\left(c^2-\frac
32\right)c_th_c^{(q)}\mbox{e}^{-c^2}\,
c\mbox{d}c \mbox{d}\theta',
\lae{apb6}
\end{equation}
\begin{equation}
u_r^{*(q)}(r)=\frac{u_r^{(q)}(r,\theta)}{\cos{\theta}}=
\frac{2}{\sqrt{\pi}}\int_{0}^{\infty}\int_{0}^{\pi}c_rc_t
h_c^{(q)}\mbox{e}^{-c^2}\,
c\mbox{d}c \mbox{d}\theta',
\lae{apb7}
\end{equation}
\begin{equation}
u_{\theta}^{*(q)}(r)=\frac{u_{\theta}^{(q)}(r,\theta)}{\sin{\theta}}=
\frac{1}{\sqrt{\pi}}\int_{0}^{\infty}\int_{0}^{\pi}
c_t^3h_s^{(q)}\mbox{e}^{-c^2}\,
c\mbox{d}c \mbox{d}\theta',
\lae{apb8}
\end{equation}
\begin{equation}
q_r^{*(q)}(r)=\frac{q_r^{(q)}(r,\theta)}{\cos{\theta}}=
\frac{2}{\sqrt{\pi}}\int_{0}^{\infty}\int_{0}^{\pi}c_rc_t\left(c^2-\frac
52\right)h_c^{(q)}\mbox{e}^{-c^2}\, c\mbox{d}c \mbox{d}\theta',
\lae{apb9}
\end{equation}
\begin{equation}
q_{\theta}^{*(q)}(r)=\frac{q_{\theta}^{(q)}(r,\theta)}{\sin{\theta}}=
\frac{1}{\sqrt{\pi}}\int_{0}^{\infty}\int_{0}^{\pi}c_t^3\left(c^2-\frac
52\right)h_s^{(q)}\mbox{e}^{-c^2}\,
c\mbox{d}c\mbox{d}\theta'.
\lae{apb10}
\end{equation}

The representation (\ref{nm1}) is compatible with the CL-model of the gas-surface
interaction so that 
\begin{equation}
\hat{A}h^{-(q)}=\cos{\theta}\hat{A}_r\hat{A}_t^{(0)}h_c^{-(q)} +
\sin{\theta}\cos{\phi'} \hat{A}_r\hat{A}_t^{(1)}h_s^{-(q)},
\lae{nm1a}
\end{equation}
where 
\[
\hat{A}_r\xi=\frac{2}{\alpha_n}\int_{0}^{\infty}
c_r'\exp{\left[-\frac{(1-\alpha_n)c_r^2+c_r'^2}{\alpha_n}\right]}
\]
\begin{equation}
\hskip2cm
\times
I_0\left(\frac{2\sqrt{1-\alpha_n}c_rc_r'}{\alpha_n}\right)\xi(-c_r',c_t')\,\mbox{d}c_r',
\lae{nm2}
\end{equation}
\[
\hat{A}_t^{(i)}\xi=\frac{2}{\alpha_t(2-\alpha_t)}\int_{0}^{\infty}c_t'^{(i+1)}
\exp{\left[-\frac{(1-\alpha_t)^2c_t^2+c_t'^2}
{\alpha_t(2-\alpha_t)}\right]}
\]
\begin{equation}
\hskip2cm
\times I_i\left[\frac{2(1-\alpha_t)c_tc_t'}{\alpha_t(2-\alpha_t)}\right]
\xi(c_r',c_t')\, \mbox{d}c_t'.
\lae{nm3}
\end{equation}
$I_i$ ($i$=0, 1) is the modified Bessel function of first kind and
$i$-th order.

Therefore, the substitution of the representation (\ref{nm1}) into the corresponding boundary condition (\ref{bc12c}) 
for the thermodynamic force $X_q$ leads to the following boundary conditions
\begin{equation}
h_c^{+(q)}=\hat{A}_r\hat{A}_t^{(0)}h_c^{-(q)}+\alpha_n(c_r^2-1)+\alpha_t(2-\alpha_t)(c_t^2-1),
\label{nm4}
\end{equation}
\begin{equation}
h_s^{+(q)}=\frac {1}{c_t}\hat{A}_r\hat{A}_t^{(1)}h_s^{-(q)}.
\lae{nm5}
\end{equation}

Far from the sphere, the representation (\ref{nm1}) leads to the following
asymptotic behaviors
\begin{equation}
\lim_{r\rightarrow \infty} h_c^{(q)}=0,
\lae{nm5a}
\end{equation}
\begin{equation}
\lim_{r\rightarrow \infty} h_s^{(q)}=0.
\lae{nm5a1}
\end{equation}

After some algebraic manipulation, the dimensionless radiometric force on the sphere
 given in (\ref{ke15}) is rewritten as
\begin{equation}
F_q=-\frac 13\left[\Pi_{rr}^{*(q)}(r_0)-2\Pi_{r\theta}^{*(q)}(r_0)\right],
\lae{nm5a2}
\end{equation}
where 
\begin{equation}
\Pi_{rr}^{*(q)}(r_0)=\frac{4}{\sqrt{\pi}}\int_{-\infty}^{\infty}\int_{-\infty}^{\infty}
c_r^2c_th_c^{(q)}(r_0,c_r,c_t)\mbox{e}^{-c^2}\,
\mbox{d}c_r\mbox{d}c_{t} ,
\lae{apb170}
\end{equation}
\begin{equation}
\Pi_{r\theta}^{*(q)}(r_0)=
\frac{2}{\sqrt{\pi}}
\int_{-\infty}^{\infty}\int_{-\infty}^{\infty}
c_rc_t^3h_s^{(q)}(r_0,c_r,c_t)\mbox{e}^{-c^2}\,
\mbox{d}c_r\mbox{d}c_{t},
\lae{apb17}
\end{equation}
with $\Pi_{rr}^{*(q)}(r_0)\cos{\theta}$ and
$\Pi_{r\theta}^{*(q)}(r_0)\sin{\theta}$ denoting the dimensionless normal
and tangential stress on the sphere. 

The system of kinetic equations (\ref{nm1a1})-(\ref{nm1a2}) subject to the boundary
conditions (\ref{nm4})-(\ref{nm5}) and
asymptotic behaviors (\ref{nm5a})-(\ref{nm5a1}) was solved numerically via the discrete velocity
 method with an accuracy of 0.1\% for the moments of the perturbation
functions at the boundary. Details regarding the discrete velocity method
can be found in the literature, see e.g. the book by Sharipov \cite{Sha02B}. The
Gaussian quadrature was used to discretize the molecular velocity and
calculate the moments of the perturbation function. The numerical values of
the nodes and weights as well as the technique to calculate them are
described in Ref. \cite{Kry02}. Moreover, a finite difference
scheme was used to approximate the derivatives which appear in the kinetic equation.
The accuracy was estimated by varying the grid parameters $N_r$, $N_{c}$ and $N_{\theta}$ corresponding to
the number of nodes in the radial coordinate $r$, molecular
speed $c$ and angle $\theta'$, as well as the maximum value of the radial
coordinate, denoted here as $r_{max}$, which defines the gas flow domain.
For $\alpha_n$=0.8, 0.9 and 1, the values of these parameters were $N_c$ and $N_{\theta}$ fixed at 20 and
1400, respectively, while $N_r$ varied according to the distance $r_{max}$ so
that the increment $\Delta r \sim 10^{-3}$. For $\alpha_n$=0.1 and 0.5 the
grid parameters were refined so that $N_c$ and $N_{\theta}$ were fixed at 25 and 1800, respectively,
while $N_r$ varied so that the increment $\Delta r \sim 10^{-4}$. The maximum
radial coordinate $r_{max}$ varied from 10 to 100 when the rarefaction parameter varied from 0.01 to 10.
The reciprocal relation (\ref{AB1}) was verified within the relative numerical error
of 0.1\% for all the values of rarefaction parameter and accommodation coefficients considered
in the calculations.

\section{Results and discussion}

\subsection{Radiometric force}

Firstly, the results obtained in the present work were compared to those
given in Ref. \cite{Tak08} for diffuse scattering, i.e. $\alpha_t$=1
and $\alpha_n$=1. Figure \ref{fig2}
shows the comparison, in which the dimensionless force $-F_q$ is plotted as
function of the rarefaction parameter $\delta$. In Ref. \cite{Tak08}, the full Boltzmann equation was solved
numerically via a finite-difference scheme method and the similarity
solution proposed in Ref. \cite{Son32}. In the present work a similar
approach was employed for solving the kinetic model to the Boltzmann equation. As one can see from Figure
\ref{fig2}, the results are in a good agreement over the whole range of
rarefaction parameter considered. The advantage of using the model kinetic equation 
instead of the full Boltzmann equation concerns the
computational effort. As it is known, in spite of the great computational
infrastructure currently available, to solve the Boltzmann equation is still
a difficult task so that the use of kinetic models plays an important role in the
solution of problems involving rarefied gas flows. 
According to Figure \ref{fig2}, as the rarefaction parameter increases the magnitude of the
radiometric force $F_q$ decreases. Note that the force is always negative,
which means a force in the $z$-direction, from the hotter to the colder side of the
sphere. 

In the free molecular regime the effects from intermolecular collisions are negligible 
and, consequently, the Knudsen mechanism is the only responsible for the appearance of the 
radiometric force on the sphere in the same direction of the gas stream
velocity ${\bf U}_{\infty}$. Since the radiometric force is due to the
momentum transfer from gas molecules to the surface, the magnitude of the
force is larger in the free molecular regime where there is no momentum
transfer due to molecular collisions. In fact, the larger the gas rarefaction
the stronger the influence of intermolecular collisions on the net momentum
transfer from gas molecules to the surface and the smaller the radiometric force on the
sphere. This behavior is also valid for other values of accommodation
coefficients. The analytic expression for the force on the
sphere in the free molecular regime is given in (\ref{new1}). 

As the rarefaction parameter increases, the
intermolecular collisions lead to the so called thermal creep 
phenomenon, which means a gas flow in the thin layer around the sphere, 
from the colder to the hotter side of the sphere, and induces a 
force on the sphere in the direction of the gas stream velocity ${\bf U}_{\infty}$.
 In the framework of the continuum equations of fluid
mechanics, since the thermal creep appears only in an approximation of first
order in the Knudsen number, the radiometric force is predicted only with the use of the 
thermal slip boundary condition and its expression is given in (\ref{new2}). Note that the force tends to zero when $\delta \rightarrow \infty$,
which means that the thermal creep flow vanishes in this limit of rarefaction parameter.
Table \ref{tab1} presents the radiometric force on the sphere obtained from
the solution of the kinetic equation as function of
the rarefaction parameter and accommodation coefficients. The range of gas
rarefaction considered in the calculations covers the free molecular,
transition and continuum regimes. The limit
solutions obtained analytically in the free molecular and continuum regimes are also presented in
Table \ref{tab1} for comparison. Since the force given in (\ref{new2}) depends on the thermal
slip coefficient, the data available in Ref. \cite{Sha84} were used.
Nonetheless, the thermal slip coefficient is sensitive to both accommodation coefficients
and it was not possible to find the values of this coefficient for all the
sets of accommodation coefficients considered in Table \ref{tab1}. 
According to Table \ref{tab1}, for diffuse scattering and $\delta$=10, 
the relative difference between the resuls obtained from the numerical
solution of the kinetic equation and that obtained from the analytic
expression (\ref{new2}) is less than 10\%. Thus, for $\delta > 10$, the
expression (\ref{new2}) can be used because this relative difference is
within 10\% when diffuse scattering is assumed. However, it is important to point out that in the vicinity of the boundary other
non-equilibrium phenomena of second order in the Knudsen number arise and influence the magnitude of the 
radiometric force, e.g. the so called thermal stress slip
flow firstly introduced by Sone \cite{Son04}. For particles with high
thermal conductivity related to that of the gas, this effect is dominant
and, as pointed out in our previous work \cite{Sha130}, it 
can explain the appearance of the negative thermophoresis in the continuum regime.

In order to analyse the influence of the NEAC and TMAC on the radiometric 
force, some numerical results are presented in Table \ref{tab1} for 
$\alpha_n$=0.1, 0.5, 0.8, 0.9 and 1, and $\alpha_t$=0.5, 0.8, 0.9 and 1.
 These values of accommodation coefficients were chosen because, in practice, the coefficients vary in the
ranges $0.1 \le \alpha_n \le 1$ and $0.6 \le \alpha_t \le 1$ for some gases,
see e.g. Ref. \cite{Sha113}. 

It is worth noting that, for all the values of rarefaction parameter
considered in the tabulated results, the reciprocity relation (\ref{AB1}) is
fullfilled for all the sets of accommodation coefficients. The relative
difference between the radiometric force given in Table \ref{tab1} and those values
obtained from Table \ref{tab3} is within 0.1\%. 

According to Table \ref{tab1}, the influence of the TMAC on the force when $\delta$=0.01
is negligible and the results tend to those predicted by the expression (\ref{new1}) in the free molecular
limit. In case of $\delta$=0.1 one can also say that the influence of the
TMAC is negligigle because the maximum deviation from the 
corresponding value of the force in case of diffuse scattering is less than
0.1\% when $\alpha_t$ varies from 1 to 0.5, but it is around 90\% when $\alpha_n$ 
varies from 1 to 0.1. As the rarefaction parameter increases, the influence of the TMAC on the
force increases, while the influence of the NEAC decreases. For instance,
for $\delta$=1, the results presented in Table \ref{tab1} show that the maximum difference from
 diffuse scattering is around 3\% when $\alpha_t$
 varies from 1 to 0.5, and 85\% when $\alpha_n$ varies from 1 to 0.1. For
$\delta$=10, the maximum difference from diffuse scattering is around 22\% when $\alpha_t$
 varies from 1 to 0.5, and 79\% when $\alpha_n$ varies from 1 to 0.1. The
comparison with the results obtained from (\ref{new2})
shows that the relative difference between the numeric and analytic results
for $\delta$=10 is larger than 10\% for the sets of accommodation
coefficients considered. As pointed out previously, this difference can be explained by the existence of other
non-equilibrium phenomena of second order in the Knudsen number whose
contribution were not considered in the analytic procedure to obtain
(\ref{new2}). 

Regarding the qualitative behavior of the force on the accommodation
coefficients, the results show us that for fixed values of the TMAC the force increases by
increasing the NEAC at arbitrary values of the gas rarefaction. Nonetheless, 
for fixed values of the NEAC, Table \ref{tab1} shows that the force decreases by increasing the
TMAC when $\delta$=1 and 10 in the range $0.5 \le \alpha_n \le 1$, while the opposite behavior
 occurs when $\alpha_n$=0.1. Moreover, note that in case of $\delta$=0.1, although the influence of the TMAC on the force
is negligible, the force always increases by increasing the TMAC. This behavior can be better explained by
visualizing the profiles of the normal and tangential stress on the sphere
as functions of the TMAC and fixed values of the NEAC. According to (\ref{nm5a2}), the
force on the sphere has the contribution of the normal and tangential stress on the spherical 
surface, which are given by the dimensionless quantities $\Pi_{rr}^{*(q)}(r_0)$ and
$\Pi_{r\theta}^{*(q)}(r_0)$. Figure \ref{fig2a} shows the profiles of $\Pi_{rr}^{*(q)}(r_0)$ and
$\Pi_{r\theta}^{*(q)}(r_0)$ as functions of the TMAC for $\delta$=0.1, 1 and
10, respectively, by considering fixed values of the NEAC, namely
$\alpha_n$=0.1, 0.3, 0.5, 0.8 and 1. As one can see
from Figure \ref{fig2a}, an increase of the TMAC leads to a small 
increase of the normal stress $\Pi_{rr}^{*(q)}$ on the sphere when
$\delta$=0.1 and 1 for all the values of the NEAC, while a decrease occurs when $\delta$=10. 
On the other hand, an increase of the 
TMAC can lead either to an increase or a decrease of the tangential stress
on the sphere as the rarefaction parameter and the NEAC vary. For instance,
when $\delta$=1 and $\alpha_n$=0.1, the tangential stress on the sphere achieves a maximum and
then decreases as the TMAC varies from 0.1 to 1. Moreover, note that in this
case, the tangential stress on the sphere is negative. Thus, the behavior of
both the normal and tangential stress on the sphere as the NEAC and TMAC vary explain the
qualitative behavior of the force given in Table \ref{tab1} when $\delta$=1
and $\alpha_n$=0.1. A similar analysis can be done for other values of
rarefaction parameter and accommodation coefficients.

\subsection{Flow fields}

The flow fields, i.e. the macroscopic characteristics of the gas flow around
the sphere corresponding to bulk velocity and heat flux in the
$z$-direction, density and temperature deviations from equilibrium, are
quantities which depend on the accommodation coefficients. To analyse
the dependence of the flow fields on the accommodation coefficients, 
the profiles of the radial and polar components of the bulk velocity 
and heat flux as well as the density and temperature deviations from equilibrium, as
functions of the distance $r/\delta$ are given in Figures
\ref{fig3}-\ref{fig10} for $\delta$=0.001, 0.1, 1 and 10. Note that, according to
the definitions given in (\ref{sp5}) and (\ref{a7}), the dimensionless
distance $r/\delta$ corresponds to the ratio of the dimensional radial coordinate
$r'$ to the radius of the sphere $R_0$. The dependence on the
TMAC is showed in Figures \ref{fig3}, \ref{fig5}, \ref{fig7} and \ref{fig9}, with the
NEAC fixed at $\alpha_n$=1. The dependence on the NEAC is showed in Figures
\ref{fig4}, \ref{fig6}, \ref{fig8} and \ref{fig10}, with the TMAC fixed at $\alpha_t$=1.
As one can see from these figures, the dependence of the flow fields on the
NEAC is stronger than that corresponding to the TMAC. In some situations
even the qualitative behavior of the gas flow is different from that
corresponding to diffuse scattering, see e.g. the radial and polar
components of the bulk velocity given in Figures \ref{fig4}, \ref{fig6} and \ref{fig8}
for $\delta$=0.001, 0.1 and 1, respectively, and $\alpha_t$=1. According to
Figures \ref{fig4}, \ref{fig6} and \ref{fig8}, in case of diffuse
scattering, the temperature
distribution on the sphere leads to a gas flow towards the positive $z$-direction when
$\delta$=0.1 and 1, and the the absence of motion when
$\delta$=0.001. However, the qualitative behavior of the gas flow can be
totally different when $\alpha_n \ne$ 1. For instance, from Figure \ref{fig4}, one can see
that, actually, there is gas motion around the sphere in the free molecular regime
when $\alpha_t$=1 and $\alpha_n \ne 1$. Moreover, such a flow is towards the negative
$z$-direction. From Figure \ref{fig6}, corresponding to $\delta$=0.1, one can also see that the gas flow is
in the negative $z$-direction when $\alpha_n$=0.1. Nonetheless, when $\alpha_n$=0.5
and 0.8 the gas flow changes direction in the layer adjacent to the sphere
so that a clockwise vortex appears. For $\delta$=1, Figure \ref{fig8} shows a similar behavior 
when $\alpha_n$=0.1. For a better visualization, the speed contour and velocity streamlines 
are given in Figures \ref{mapdel01}-\ref{mapdel10} for $\delta$=0.1, 1 and
10, $\alpha_t$=1, $\alpha_n$=0.1 and 0.5. These figures clearly show that
the flow pattern around the sphere can drastically change as the
 NEAC varies in the free molecular and transition regime. For fixed values of
the NEAC a change in the flow pattern is also observed when the rarefaction
parameter changes. For instance, when $\delta$=10 the gas flows towards the positive
$z$-direction when $\alpha_n$=0.1. However, a clockwise vortex appears in the vicinity 
of the boundary when $\delta$=1, and the transition of the flow pattern from
the positive to the negative $z$-direction is complete when $\delta$=0.1. This kind
of change in the flow pattern as the accommodation coefficients vary in the
free molecular and transition regimes was already
observed in Ref. \cite{Kos09} for a planar geometry. 

From Figures \ref{fig3}, \ref{fig5}, \ref{fig7} and \ref{fig9} one can see
that the temperature and density deviations from equilibrium did not depend on
the TMAC in the whole range of the gas rarefaction. According to these figures, while
the temperature deviation increases near the sphere, the density deviation
decreases. On the other hand, Figures
\ref{fig4}, \ref{fig6}, \ref{fig8} and \ref{fig10} show a dependence of these quantities on the NEAC.
For $\delta$=10 the temperature deviation increases while the density
deviation decreases near the sphere. However, for other values of the gas
rarefaction, only que qualitative behavior of the temperature deviation is
the same. As one can see, when $\delta$=0.001, 0.1 and 1, the density
deviation decreases near the sphere when $\alpha_n \ge 0.5$, but when
$\alpha_n$=0.1 there is an increase of the density deviation in the thin
layer adjacent to the sphere.


\subsection{Comparison with experiment}

In practice, the thermal polarization on the spherical particle is
measured, i.e. the temperature drop between the ends of the sphere diameter
parallel to the gas stream velocity ${\bf U}_{\infty}$. Here, this
temperature drop is obtained from (\ref{la4}) as
\begin{equation}
\Delta T'=2T_0 \tau_{s0},
\lae{tp1}
\end{equation}		
where $\tau_{s0}$=$X_q$. 

The condition (\ref{la2}) allows us to
relate the thermodynamic forces $X_q$ and $X_u$ so that 
\begin{equation}
X_q=-q_r^{*(u)}(r_0)\left[q_r^{*(q)}(r_0)+\frac{15}{8}\frac{\Lambda}{\delta}\right]^{-1}X_u,
\lae{tp2}
\end{equation} 
where the values for the radial components of the heat flux at the surface,
$q_r^{*(q)}(r_0)$ and $q_r^{*(u)}(r_0)$, due to each thermodynamic force are given in
Tables \ref{tab2} and \ref{tab3} for some values of rarefaction parameter
and accommodation coefficients. Note that, in the free molecular regime,
while $q_r^{*(u)}(r_0)$ does not depend on the TMAC, $q_r^{*(q)}(r_0)$ depends on both
accommodation coefficients. Moreover, in the limit $\Lambda \rightarrow \infty$
corresponding to a sphere with high thermal conductivity related to that of
the carrier gas, $\Delta T \rightarrow 0$ for arbitrary values of
rarefaction parameter so that the thermal polarization
of the spherical particle can be neglected. 

For convenience, from (\ref{tp1}) and
(\ref{tp2}), the following dimensionless temperature drop is introduced as
\begin{equation}
\Delta T=\frac{\Delta T'}{T_0X_u}=-2q_r^{*(u)}(r_0)\left(q_r^{*(q)}(r_0)+\frac{15}{8}\frac{\Lambda}{\delta}\right)^{-1}.
\lae{tp3}
\end{equation}

Figures \ref{fig14a} and \ref{fig14b} show the results for the temperature drop $\Delta T$ 
as function of the rarefaction parameter $\delta$, in the range $0 < \delta \le
10$, for a Pyrex glass sphere in helium ($\Lambda$=7.5)
 and argon ($\Lambda$=62.5) gas. The experimental data provided in Ref.
\cite{Bak05} are presented in theses figures for comparison. It is worth noting that these experimental data did
not cover the whole range of the gas rarefaction we are interested in. While
for helium the data cover the range $3 < \delta \le 10$, for
argon the range is reduced to $6 <\delta \le 10$.  
The experimental apparatus employed in Ref. \cite{Bak05} consisted of a 6 mm radius
sphere suspended by a differential thermocouple in the center of a tube 1.8 m long
and 0.12 m in diameter mounted in a thermostatically controlled box. The
pressure of the system was adjusted to vary in the range from 1 to 100 Pa,
while a gas handling system allowed to provide a volume flow within the
range from 0 to 20 cm$^3$/s corresponding to Mach number from $10^{-2}$ to
$10^{-3}$. The thermocouple signal was recorded by a set of instruments so
that the value of the temperature drop between the ends of the sphere was
measured at different values of Mach number and various values of fixed gas pressures
in the tube. The dimensionless temperature drop introduced in Ref.
\cite{Bak05}, which is denoted by $\Delta T^*$, as function of the Knudsen number,
$Kn$=$\sqrt{\pi}/(2\delta)$, is
related to the temperature drop defined in (\ref{tp3}) as 
\begin{equation}
\Delta T=\frac{\pi}{2\delta}\frac{\Delta T^*}{(\Lambda +2)}.
\end{equation}

According to the literature, see e.g. Ref. \cite{Sha84},
the TMAC of helium and argon at a glass surface supported by experiments is
close to 0.9, while no values were found for the NEAC of both gases at the
same surface. However, it is known that the NEAC of helium in clean metallic surfaces
is close to 0.1, while in case of argon is close to 0.6. Thus, the 
temperature drop was calculated for diffuse scattering and for other sets of accommodation
coefficients with the TMAC fixed at 0.9 and the NEAC varying in the range
$0.1 \le \alpha_n \le 1$. Figures \ref{fig14a} and \ref{fig14b}
show the temperature drop $\Delta T$ as function of the rarefaction
parameter for some sets of accommodation coefficients.

According to Figure
\ref{fig14a}, corresponding to helium, the numerical results are in good agreement with the
experimental data provided by Ref. \cite{Bak05} when $\alpha_t$=0.9 and $\alpha_n$=0.4.
However, the experimental data did not cover all the range of the gas
rarefaction considered in the numerical calculations. For argon, as one can
see from Figure \ref{fig14b}, a good agreement with experimenal data is observed when $\alpha_t$=0.9 and 
$\alpha_n$=0.8 and just one experimental point is closer to the theoretical
results corresponding to $\alpha_n$=0.6. Nonetheless, it is worth noting
that the range of rarefaction parameter covered by the experiment with argon
gas is smaller than that corresponding to helium gas. As expected, since the thermal
conductivity of argon is smaller than that corresponding to helium, the
thermal polarization effect of the spherical particle is stronger in helium.
Moreover, note that the temperature drop has a non-linear behavior in the rarefaction
parameter. According to Figures \ref{fig14a} and \ref{fig14b}, the thermal
polarization of the Pyrex glass sphere in both gases is stronger when the
rarefaction parameter is in the range $2 < \delta < 4$.

\section{Concluding remarks}

In the present work, the radiometric force acting on a spherical particle with
arbitrary thermal conductivity, as well as the flow fields around it, were calculated on the basis of the linearized
kinetic equation proposed by Shakhov and the Cercignani-Lampis model of
gas-surface interaction. The kinetic equation was solved by using the
discrete velocity method in a range of the gas rarefaction covering the
free molecular, transition and continuum regimes, and for some sets of
accommodation coefficients. The reciprocity relation between the cross
phenomena was obtained and verified numerically within the numerical error.
The results show a significant dependence of the force and flow fields on the TMAC and NEAC. For fixed values
of the TMAC, the force always increases by increasing the NEAC. However, for
fixed values of the NEAC, the qualitative behavior of the force depends on
the values of the TMAC and rarefaction parameter. In the free
molecular regime the radiometric force is independent of the TMAC, while in the transition and
continuum regimes this force strongly depends on both accommodation
coefficients, and it can change even qualitatively its behavior. The gas flow pattern around
the sphere is also strongly dependent on the accommodation coefficients. The
absence of gas motion around the sphere under the diffuse scattering
predicted previously is observed. However, for non-diffuse
scattering there is indeed a gas flow around the sphere. The dominant gas flow can be directed in the positive
 $z$-direction with the counterclockwise vortex in the vicinity of the
sphere. In some situatios, the flow direction is opposite, i.e. the vortex
is clockwise. 

 The results for the temperature drop between the two
diametrically opposite points on the spherical surface in the $z$-direction
was calculated and compared with experimental data for a Pyrex glass in helium and argon.
The comparison shows a good agreement between the numerical and experimental
results when appropriate values for the NEAC and TMAC are used. The resuls
obtained in the present work show the importance of the gas-surface
interaction law for the correct modelling of the physics underlying the
movement of small particles in a rarefied gas, such as aerosols in the
atmosphere and in technological applications.  									

\section*{Declaration of Competing Interest}

The authors declare that they have no competing financial interests or
personal relationships that could have appeared to influence the work
reported in this paper.

\section*{Acknowledgments}

D. Kalempa acknowledges FAPESP (Funda\c{c}\~ao de Amparo \`a Pesquisa do
Estado de S\~ao Paulo) for the support of the research, grant 2015/20650-5. 
F. Sharipov acknowledges CNPq (Conselho Nacional de Desenvolvimento
Cient\'{\i}fico e Tecnol\'ogido), grant 304831/2018-2.


\bibliographystyle{unsrt}

\biboptions{numbers,sort&compress}


\clearpage


\begin{table}
\centering
\begin{threeparttable}
\begin{tabular}{ccccccc}
 & & \multicolumn{4}{c}{$-F_q$}\\ \cline{3-7}
             & $\alpha_t$ & $\alpha_n$=0.1 & 0.5 & 0.8 & 0.9 & 1.0  \\ \hline
$\delta\rightarrow 0$\tnote{a}  & --- &0.008333 &0.04167 &0.06667 &0.07500 &0.08333 \\[0.25cm]
$\delta$=0.01& 0.5 & 0.008408 & 0.04166 & 0.06658 & 0.07489 & 0.08319  \\
             & 0.8 & 0.008444 & 0.04169 & 0.06660 & 0.07491 & 0.08321  \\
             & 0.9 & 0.008451 & 0.04169 & 0.06661 & 0.07491 & 0.08322 \\
             & 1.0 & 0.008456 & 0.04170 & 0.06661 & 0.07491 & 0.08322  \\[0.25cm]
$\delta$=0.1 & 0.5 & 0.008848 & 0.04125 & 0.06547 & 0.07354 & 0.08162  \\
             & 0.8 & 0.009114 & 0.04142 & 0.06556 & 0.07360 & 0.08164 \\
             & 0.9 & 0.009169 & 0.04145 & 0.06557 & 0.07361 & 0.08164 \\
             & 1.0 & 0.009206 & 0.04147 & 0.06559 & 0.07360 & 0.08164 \\[0.25cm]
$\delta$=1   & 0.5 & 0.008023 & 0.03170 & 0.04906 & 0.05475 & 0.06041 \\
             & 0.8 & 0.008579 & 0.03140 & 0.04812 & 0.05369 & 0.05914 \\
             & 0.9 & 0.008687 & 0.03127 & 0.04783 & 0.05336 & 0.05875 \\
             & 1.0 & 0.008750 & 0.03114 & 0.04755 & 0.05303 & 0.05838 \\[0.25cm] 
$\delta$=10  & 0.5 & 0.001076 & 0.004486 & 0.006530 & 0.007134 & 0.007705  \\
             & 0.8 & 0.001220 & 0.004008 & 0.005694 & 0.006193 & 0.006664 \\
             & 0.9 & 0.001278 & 0.003924 & 0.005542 & 0.006017 & 0.006464  \\
             & 1.0 & 0.001335 & 0.003875 & 0.005429 & 0.005884 & 0.006312 \\ [0.25cm]
$\delta$=10\tnote{b} & 0.5 & ----- &0.005405 & ----- & ----- &0.005860 \\
             & 1.0 & ----- &0.005875 & ----- & -----&0.005875 \\ \hline
\end{tabular}
\begin{tablenotes}
\footnotesize
\item[a] Eq. (\ref{new1}), free molecular regime.
\item[b] Eq. (\ref{new2}), slip flow regime.
\end{tablenotes}
\end{threeparttable}
\caption{Dimensionless radiometric force vs. rarefaction parameter and
accommodation coefficients.}
\lae{tab1}
\end{table}

\begin{table}
\centering
\begin{threeparttable}
\begin{tabular}{ccccccc}
 & & \multicolumn{4}{c}{$q_r^{*(q)}$}\\ \cline{3-7}
             & $\alpha_t$ & $\alpha_n$=0.1 & 0.5 & 0.8 & 0.9 & 1.0  \\ \hline
$\delta\rightarrow 0$\tnote{a}  & 0.5 & 0.2398 & 0.3526 & 0.4372 & 0.4654 & 0.4937  \\
             & 0.8 & 0.2990 & 0.4118 & 0.4965 & 0.5247 & 0.5529  \\
             & 0.9 & 0.3075 & 0.4203 & 0.5049 & 0.5331 & 0.5614  \\
             & 1.0 & 0.3103 & 0.4231 & 0.5078 & 0.5360 & 0.5642  \\[0.25cm]
$\delta$=0.01& 0.5 & 0.2397 & 0.3524 & 0.4369 & 0.4651 & 0.4932 \\
             & 0.8 & 0.2988 & 0.4115 & 0.4961 & 0.5242 & 0.5524   \\
             & 0.9 & 0.3073 & 0.4200 & 0.5045 & 0.5327 & 0.5608 \\
             & 1.0 & 0.3101 & 0.4228 & 0.5073 & 0.5355 & 0.5637  \\[0.25cm]
$\delta$=0.1 & 0.5 & 0.2387 & 0.3505 & 0.4341 & 0.4620 & 0.4898  \\
             & 0.8 & 0.2974 & 0.4089 & 0.4924 & 0.5202 & 0.5479 \\
             & 0.9 & 0.3057 & 0.4173 & 0.5007 & 0.5285 & 0.5562 \\
             & 1.0 & 0.3085 & 0.4200 & 0.5035 & 0.5312 & 0.5590 \\[0.25cm]
$\delta$=1   & 0.5 & 0.2287 & 0.3314 & 0.4067 & 0.4314 & 0.4560 \\
             & 0.8 & 0.2819 & 0.3826 & 0.4564 & 0.4808 & 0.5048 \\
             & 0.9 & 0.2894 & 0.3899 & 0.4634 & 0.4876 & 0.5116 \\
             & 1.0 & 0.2919 & 0.3923 & 0.4658 & 0.4899 & 0.5139  \\[0.25cm] 
$\delta$=10  & 0.5 & 0.1518 & 0.1952 & 0.2210 & 0.2286 & 0.2358  \\
             & 0.8 & 0.1733 & 0.2111 & 0.2339 & 0.2406 & 0.2469 \\
             & 0.9 & 0.1760 & 0.2131 & 0.2355 & 0.2422 & 0.2484  \\
             & 1.0 & 0.1769 & 0.2138 & 0.2361 & 0.2427 & 0.2489 \\ \hline
\end{tabular}
\begin{tablenotes}
\footnotesize
\item[a] Analytic solution in the free molecular regime obtained by
substituting (\ref{bc12c}) into (\ref{k12f}).
\end{tablenotes}
\end{threeparttable}
\caption{Dimensionless radial component of the heat flux on the spherical
surface due to the force $X_q$.}
\lae{tab2}
\end{table}

\begin{table}
\centering
\begin{threeparttable}
\begin{tabular}{ccccccc}
 & & \multicolumn{4}{c}{$q_r^{*(u)}$}\\ \cline{3-7}
             & $\alpha_t$ & $\alpha_n$=0.1 & 0.5 & 0.8 & 0.9 & 1.0  \\ \hline
$\delta\rightarrow 0$\tnote{a}  & --- & 0.02500 & 0.1250 & 0.2000 & 0.2250 & 0.2500 \\[0.25cm]
$\delta$=0.01& 0.5 & 0.02521 & 0.1250 & 0.1997 & 0.2247 & 0.2496 \\
             & 0.8 & 0.02531 & 0.1250 & 0.1998 & 0.2247 & 0.2496 \\
             & 0.9 & 0.02533 & 0.1250 & 0.1998 & 0.2247 & 0.2496  \\
             & 1.0 & 0.02535 & 0.1251 & 0.1998 & 0.2247 & 0.2497 \\[0.25cm]
$\delta$=0.1 & 0.5 & 0.02653 & 0.1237 & 0.1964 & 0.2206 & 0.2449   \\
             & 0.8 & 0.02734 & 0.1242 & 0.1966 & 0.2208 & 0.2449  \\
             & 0.9 & 0.02751 & 0.1243 & 0.1966 & 0.2208 & 0.2449  \\
             & 1.0 & 0.02762 & 0.1243 & 0.1966 & 0.2207 & 0.2449 \\[0.25cm]
$\delta$=1   & 0.5 & 0.02405 & 0.09506 & 0.1472 & 0.1644 & 0.1816   \\
             & 0.8 & 0.02574 & 0.09411 & 0.1444 & 0.1610 & 0.1775  \\
             & 0.9 & 0.02606 & 0.09375 & 0.1435 & 0.1600 & 0.1763 \\
             & 1.0 & 0.02625 & 0.09334 & 0.1427 & 0.1590 & 0.1752  \\[0.25cm] 
$\delta$=10  & 0.5 & 0.003225 & 0.01345 & 0.01961 & 0.02143 & 0.02315   \\
             & 0.8 & 0.003656 & 0.01199 & 0.01707 & 0.01859 & 0.02001  \\
             & 0.9 & 0.003832 & 0.01177 & 0.01661 & 0.01806 & 0.01941 \\
             & 1.0 & 0.004007 & 0.01163 & 0.01627 & 0.01766 & 0.01894 \\ \hline
\end{tabular}
\begin{tablenotes}
\footnotesize
\item[a] Eq. (\ref{new3}), free molecular regime.
\end{tablenotes}
\end{threeparttable}
\caption{Dimensionless radial component of the heat flux on the spherical
surface due to the force $X_u$.}
\lae{tab3}
\end{table}

\clearpage

\begin{figure}
\centering
\includegraphics[scale=1]{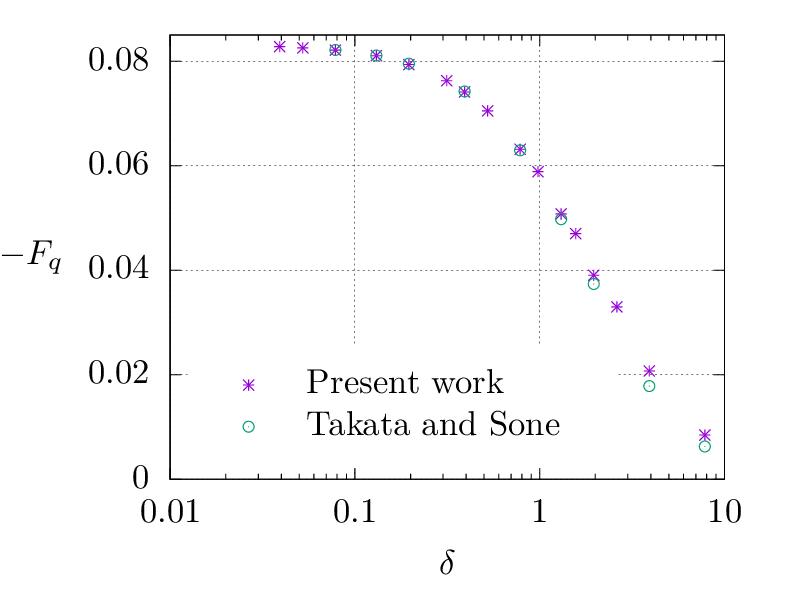}
\caption{Radiometric force versus rarefaction parameter:
comparison to the results by Takata and Sone \cite{Tak08}, diffuse scattering.}
\lae{fig2}
\end{figure}

\clearpage

\begin{figure}
\centering
\subfigure[$\delta$=0.1]{
\includegraphics[scale=0.9]{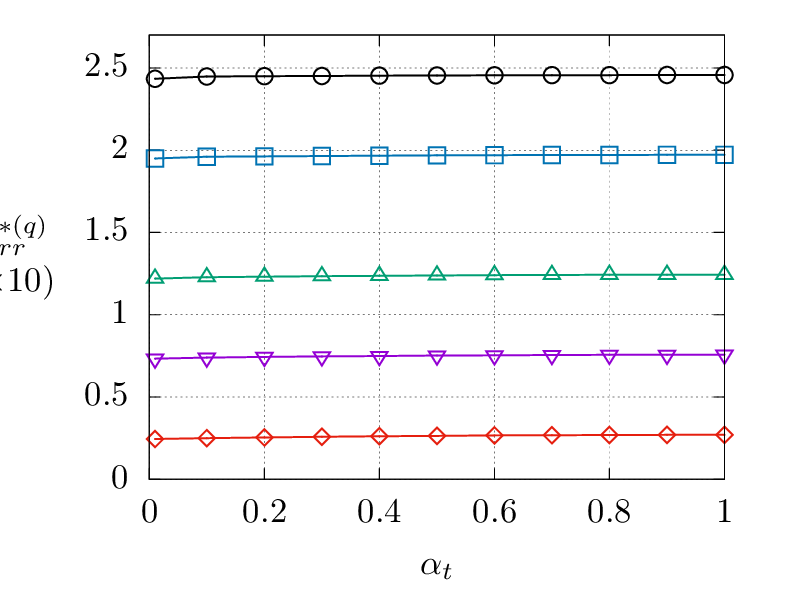}
\includegraphics[scale=0.9]{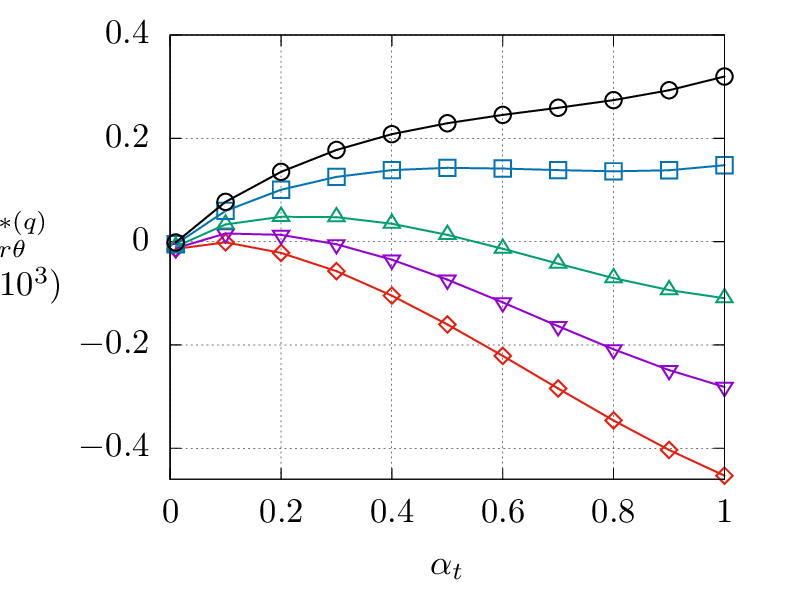}
}
\subfigure[$\delta$=1]{
\includegraphics[scale=0.9]{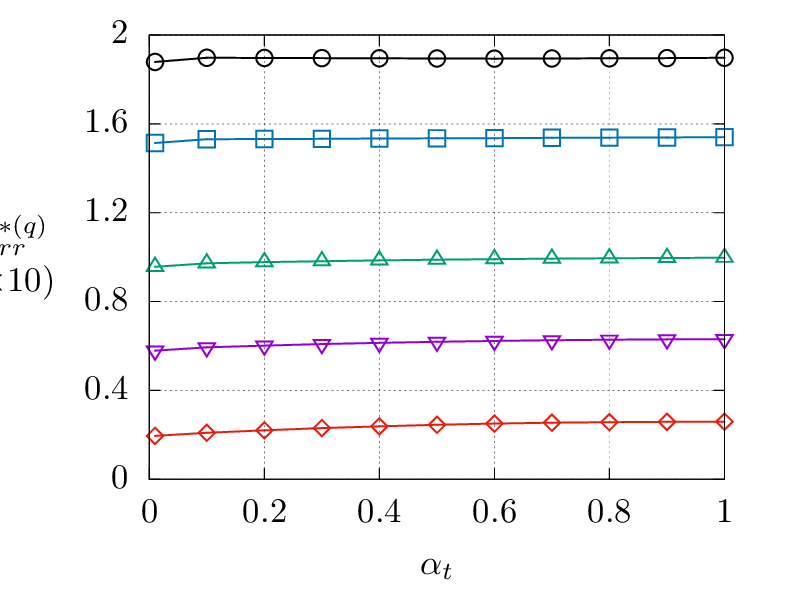}
\includegraphics[scale=0.9]{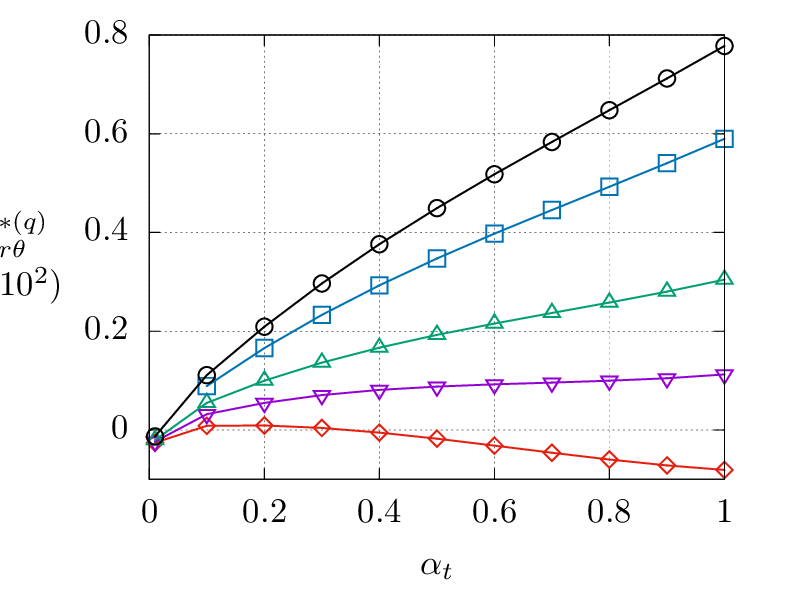}
}
\subfigure[$\delta$=10]{
\includegraphics[scale=0.9]{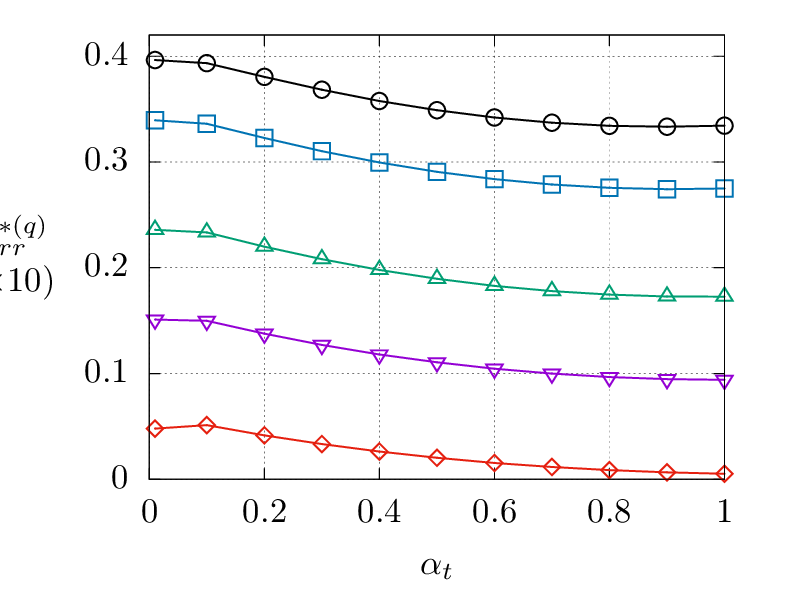}
\includegraphics[scale=0.9]{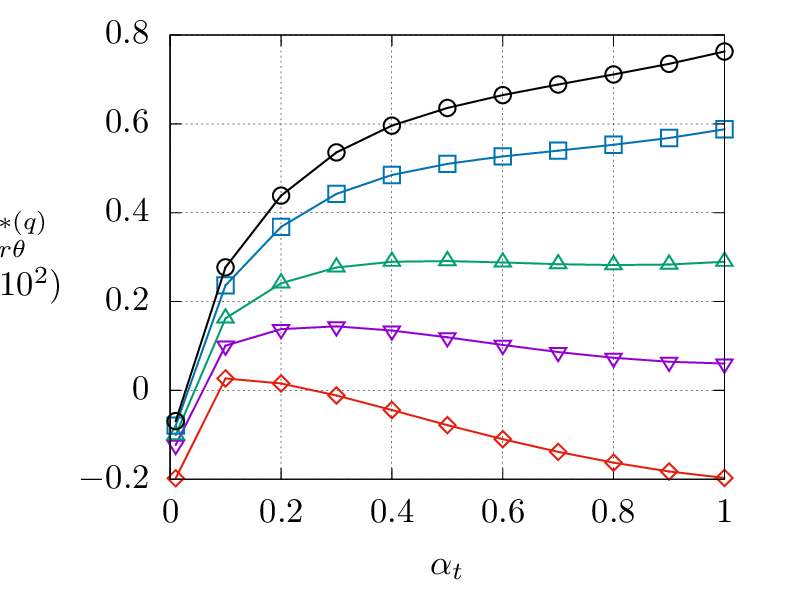}
}
\caption{Normal and tangential stress on the sphere as function of the TMAC
at fixed values of the NEAC. Symbols: $\circ$ - $\alpha_n$=1; $\square$ - 
$\alpha_n$=0.8; $\vartriangle$ - $\alpha_n$=0.5; $\triangledown$ - 
$\alpha_n$=0.3; $\lozenge$ - $\alpha_n$=0.1.} 
\lae{fig2a}
\end{figure}


\begin{figure}
\centering
\subfigure[Radial and polar components of the bulk velocity]{
\includegraphics[scale=0.9]{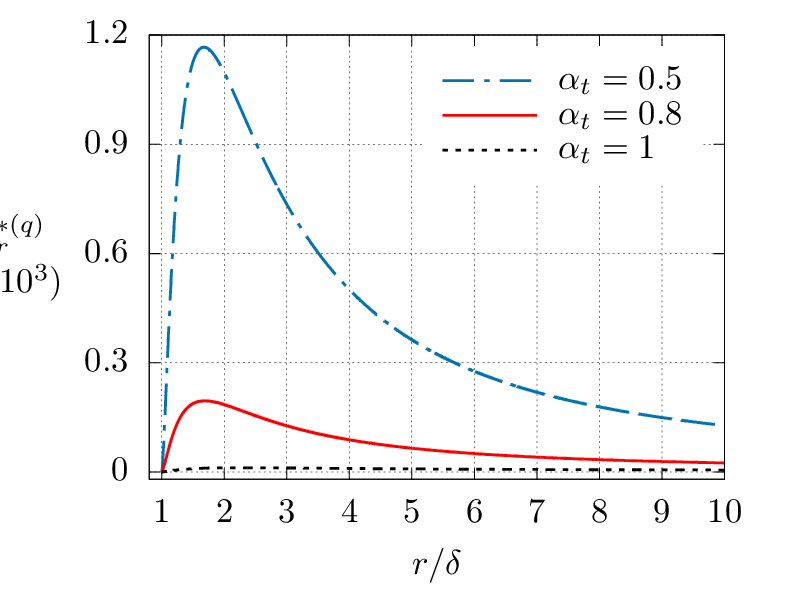}
\includegraphics[scale=0.9]{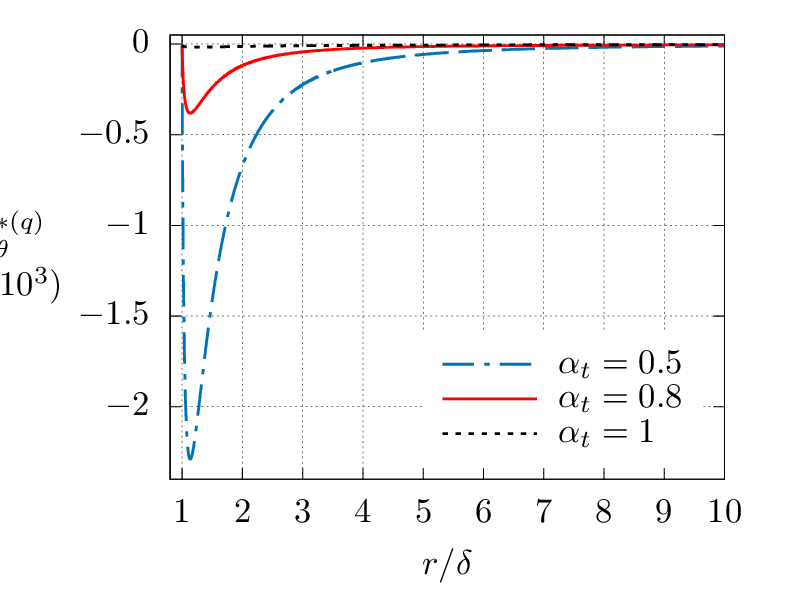}
}
\subfigure[Radial and polar components of the heat flux]{
\includegraphics[scale=0.9]{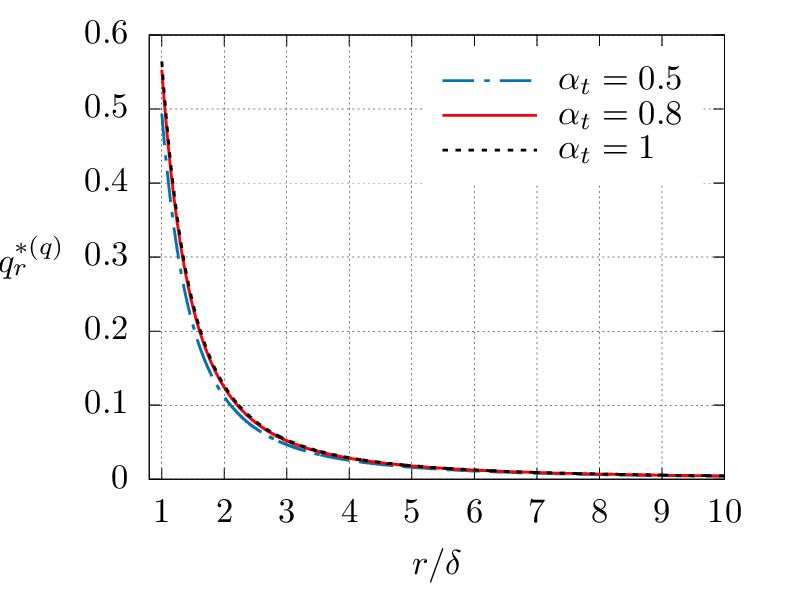}
\includegraphics[scale=0.9]{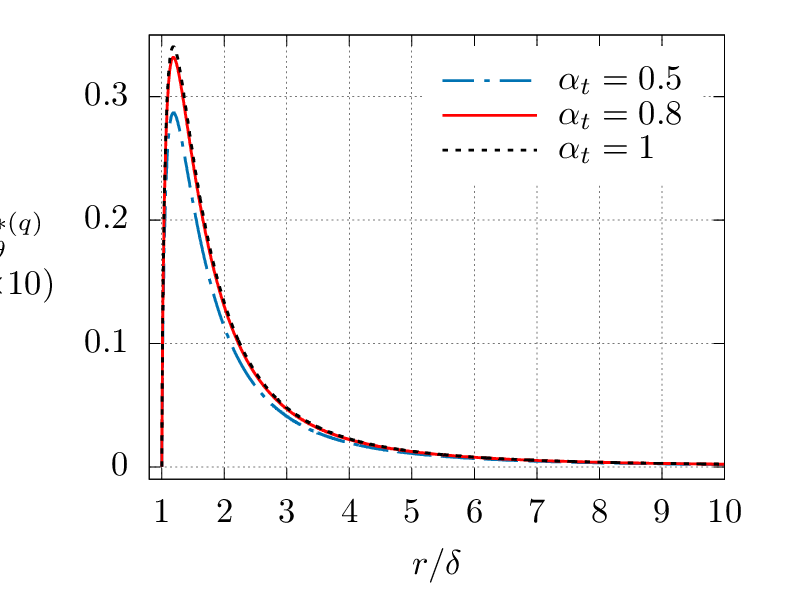}
}
\subfigure[Density and temperature deviations from equilibrium]{
\includegraphics[scale=0.9]{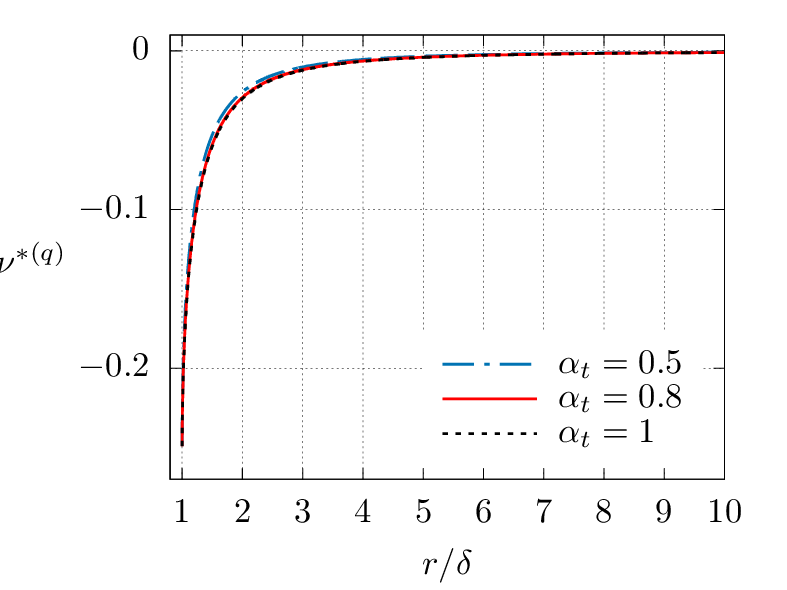}
\includegraphics[scale=0.9]{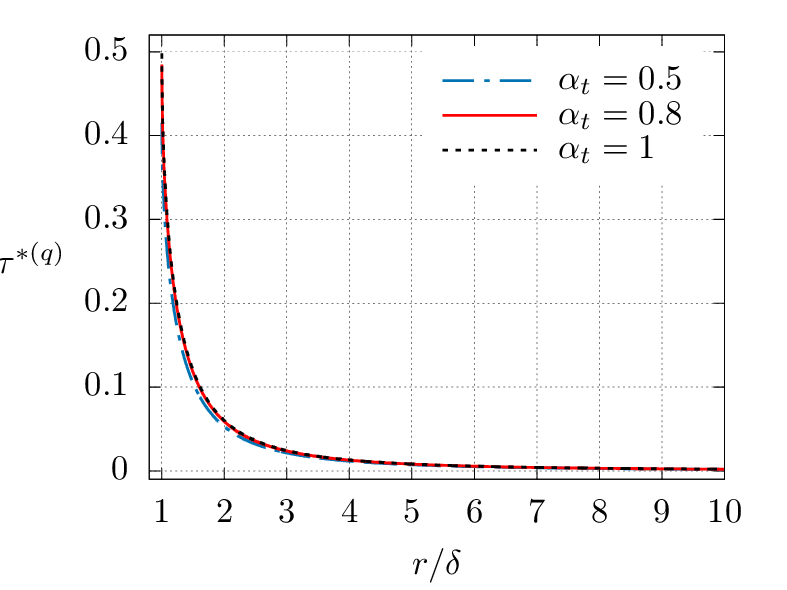}
}
\caption{Flow fields due to the thermodynamic
force $X_q$ for fixed $\alpha_n$=1 and $\delta$=0.001.}
\lae{fig3}
\end{figure}

\begin{figure}
\centering
\subfigure[Radial and polar components of the bulk velocity]{
\includegraphics[scale=0.9]{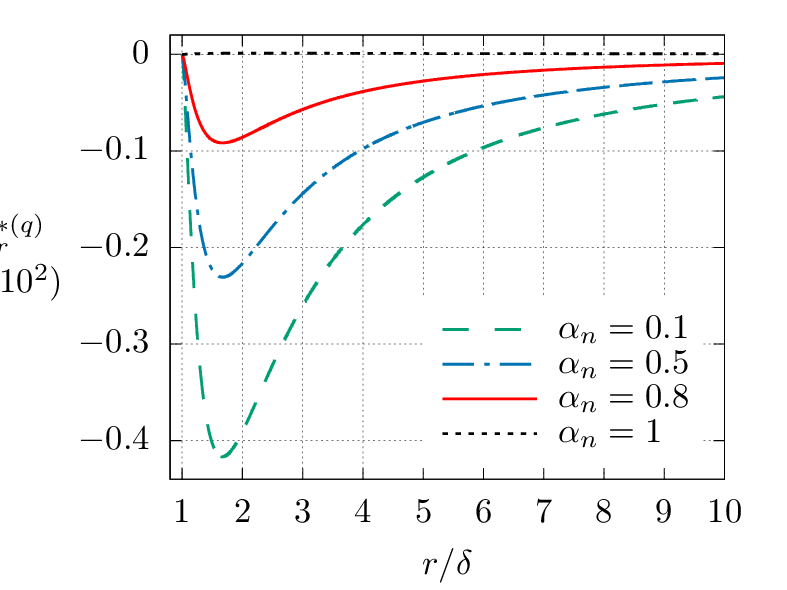}
\includegraphics[scale=0.9]{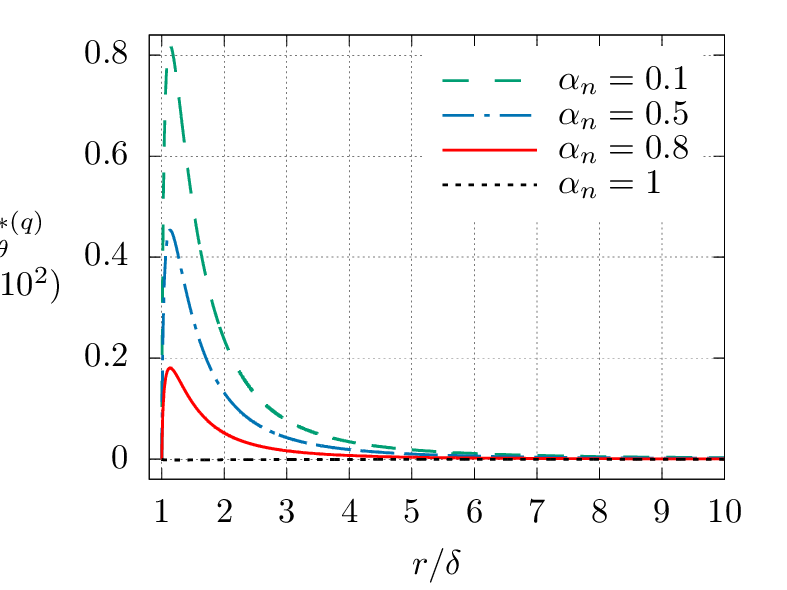}
}
\subfigure[Radial and polar components of the heat flux]{
\includegraphics[scale=0.9]{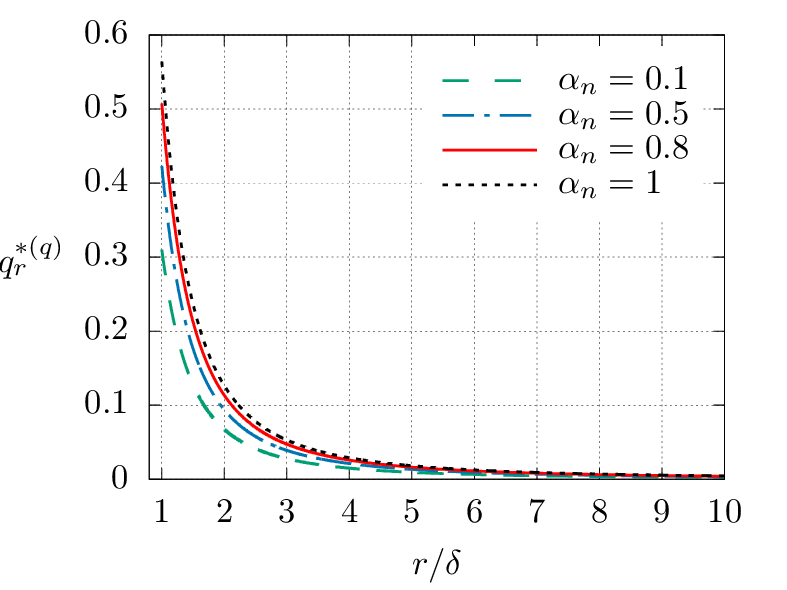}
\includegraphics[scale=0.9]{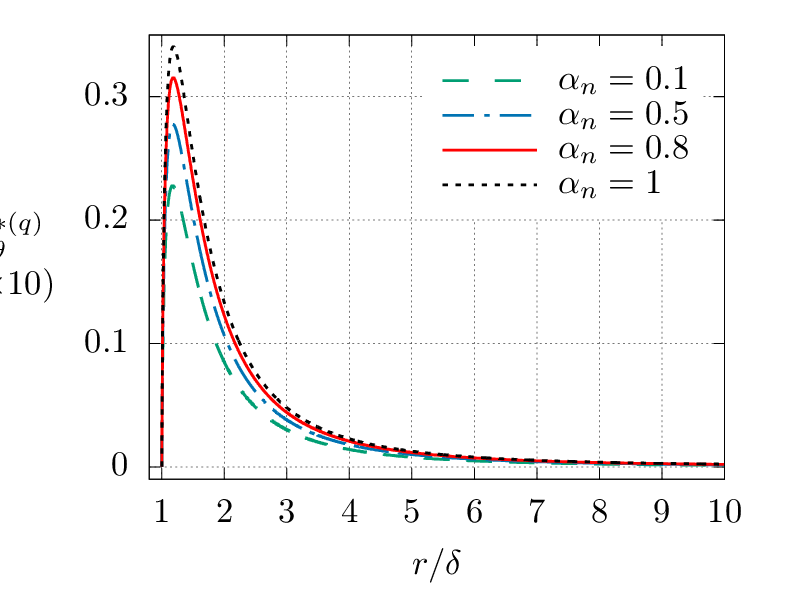}
}
\subfigure[Density and temperature deviations from equilibrium]{
\includegraphics[scale=0.9]{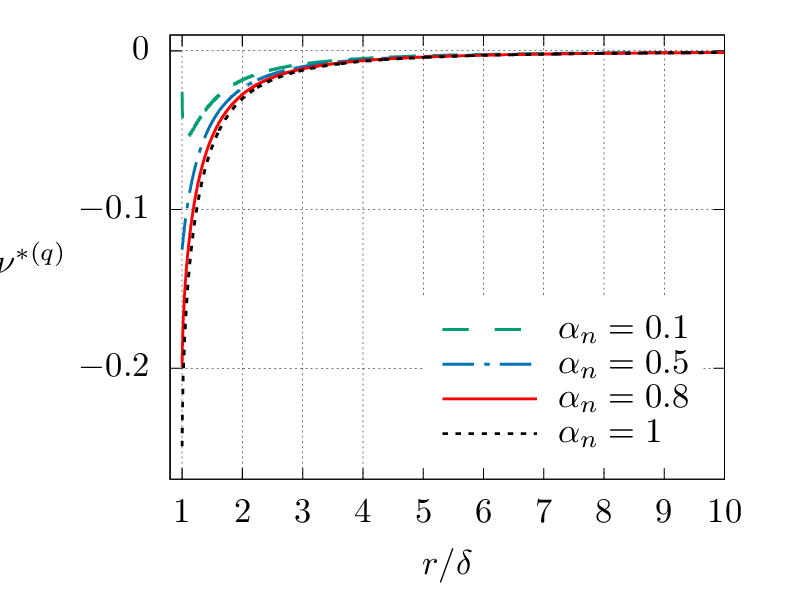}
\includegraphics[scale=0.9]{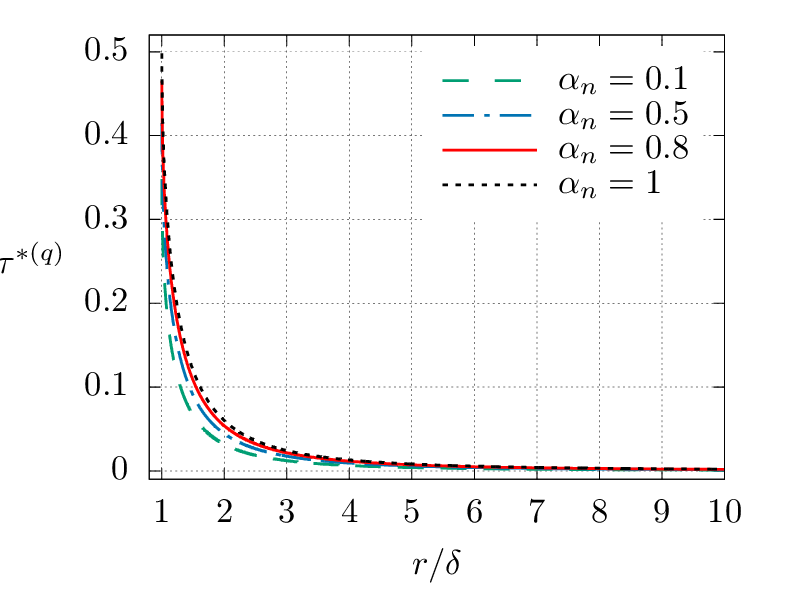}
}
\caption{Flow fields due to the thermodynamic
force $X_q$ for fixed $\alpha_t$=1 and $\delta$=0.001.}
\lae{fig4}
\end{figure}


\begin{figure}
\centering
\subfigure[Radial and polar components of the bulk velocity]{
\includegraphics[scale=0.9]{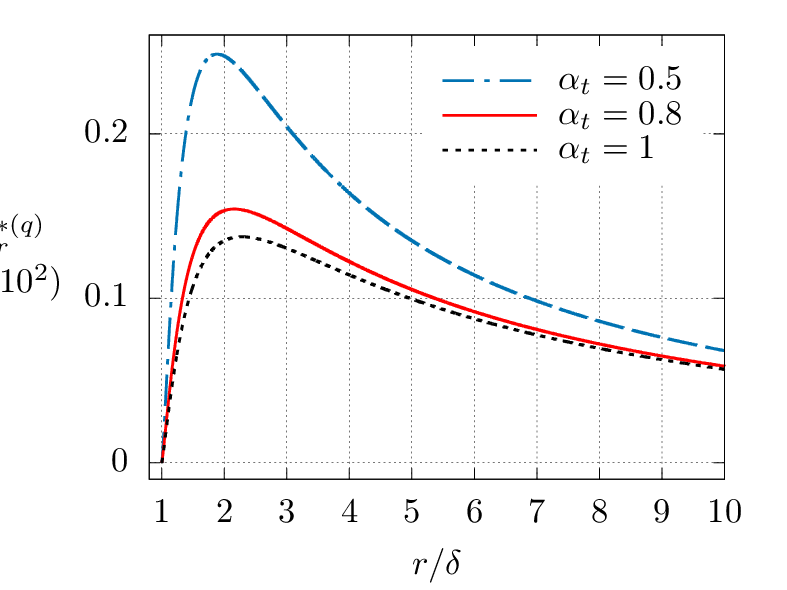}
\includegraphics[scale=0.9]{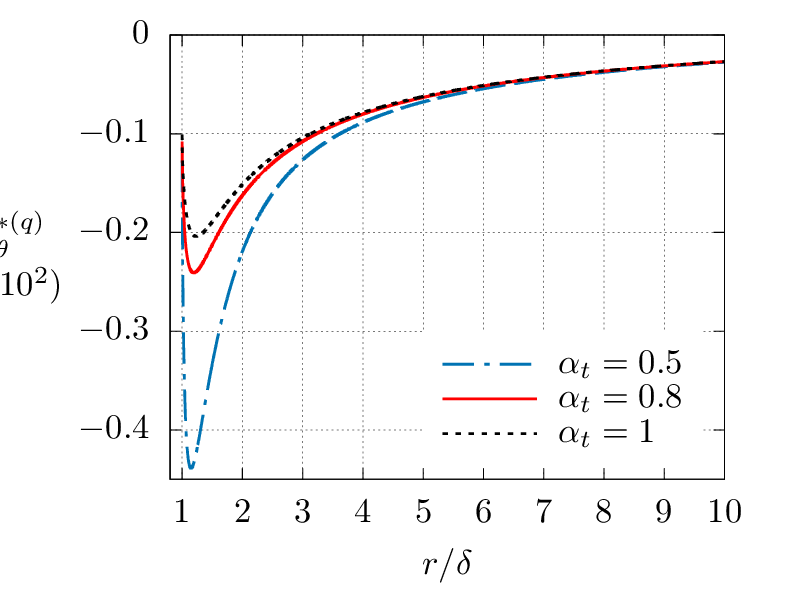}
}
\subfigure[Radial and polar components of the heat flux]{
\includegraphics[scale=0.9]{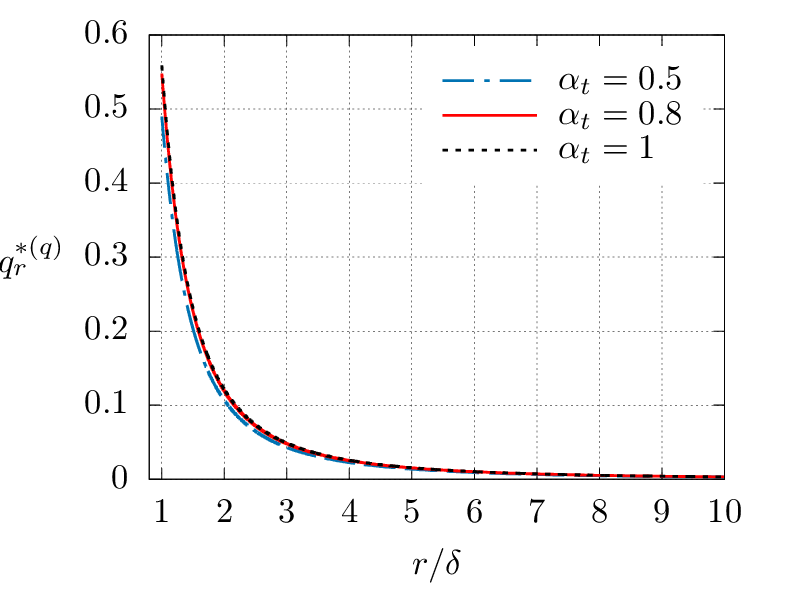}
\includegraphics[scale=0.9]{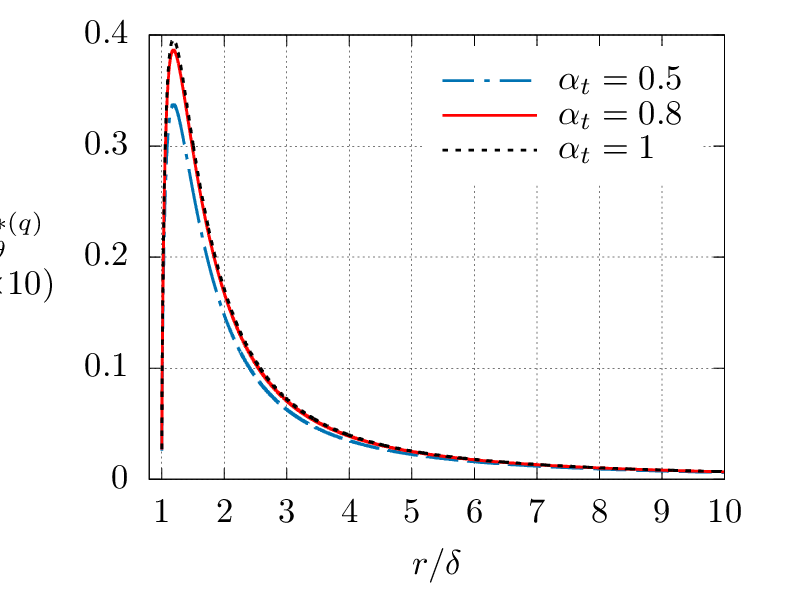}
}
\subfigure[Density and temperature deviations from equilibrium]{
\includegraphics[scale=0.9]{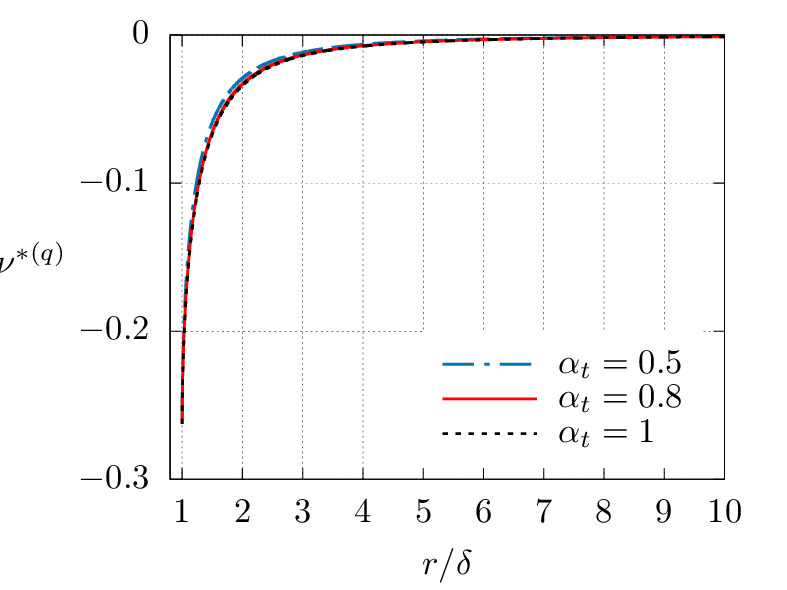}
\includegraphics[scale=0.9]{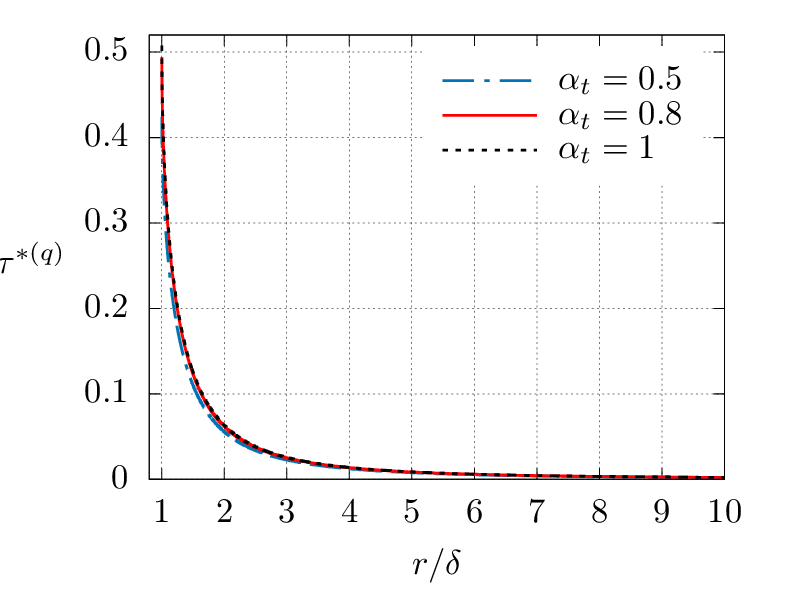}
}
\caption{Flow fields due to the thermodynamic
force $X_q$ for fixed $\alpha_n$=1 and $\delta$=0.1.}
\lae{fig5}
\end{figure}

\begin{figure}
\centering
\subfigure[Radial and polar components of the bulk velocity]{
\includegraphics[scale=0.9]{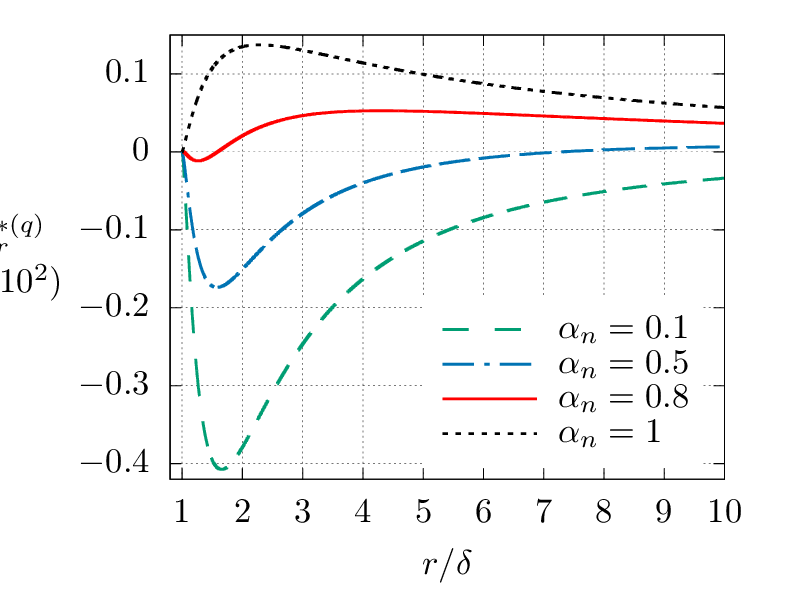}
\includegraphics[scale=0.9]{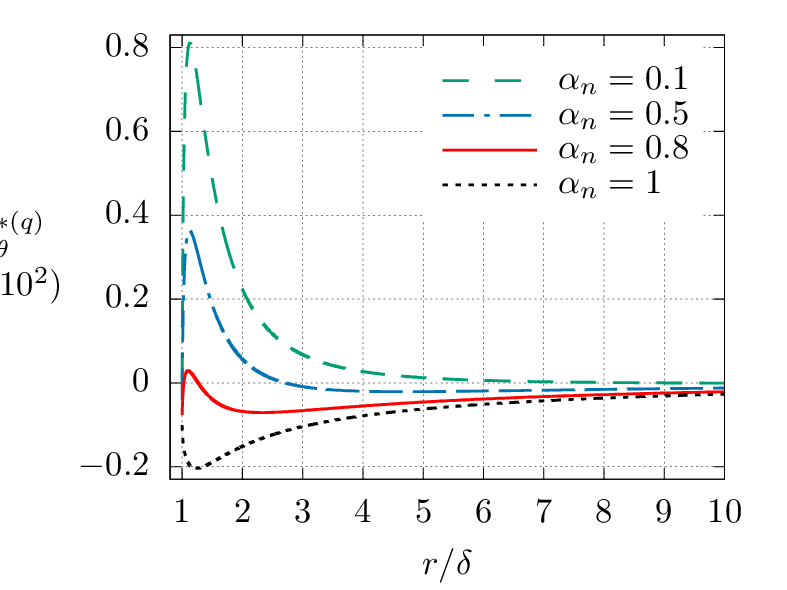}
}
\subfigure[Radial and polar components of the heat flux]{
\includegraphics[scale=0.9]{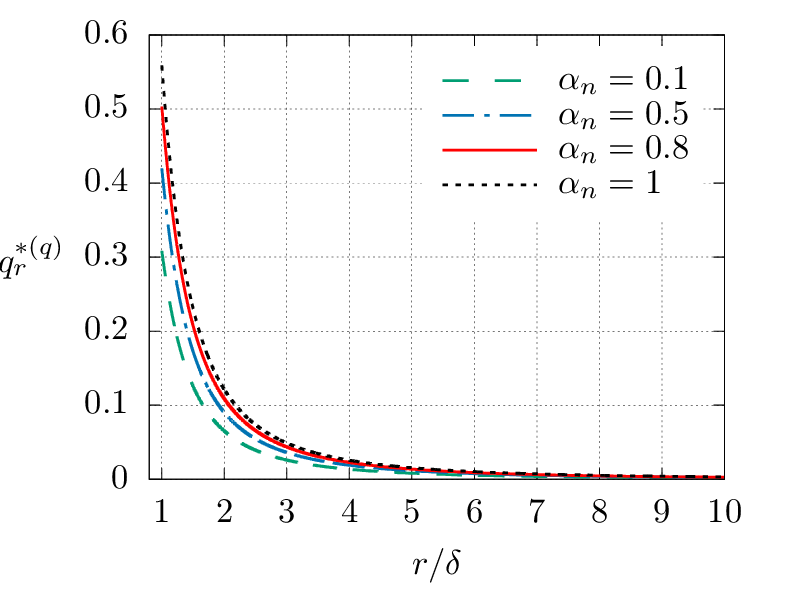}
\includegraphics[scale=0.9]{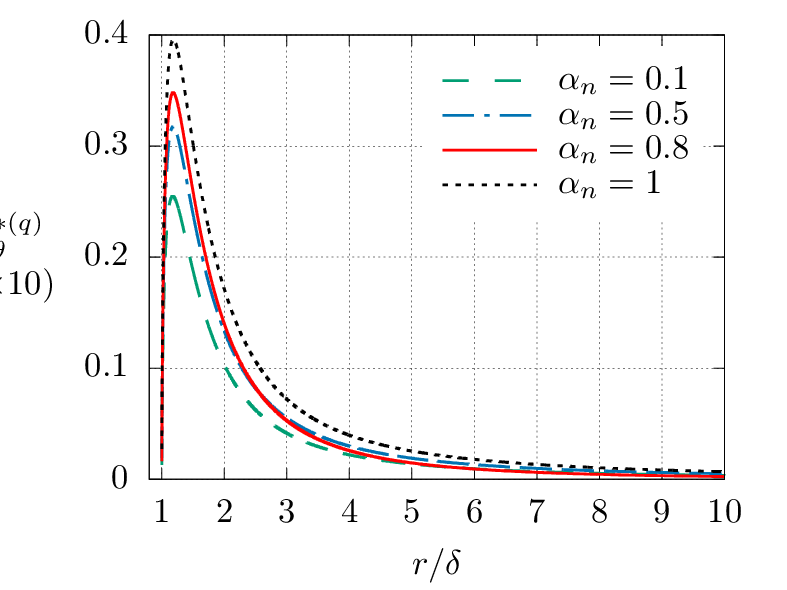}
}
\subfigure[Density and temperature deviations from equilibrium]{
\includegraphics[scale=0.9]{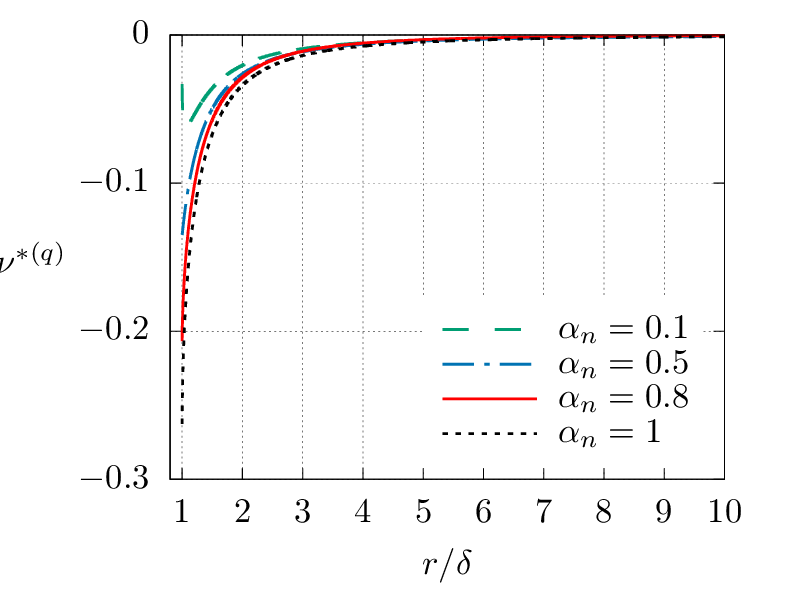}
\includegraphics[scale=0.9]{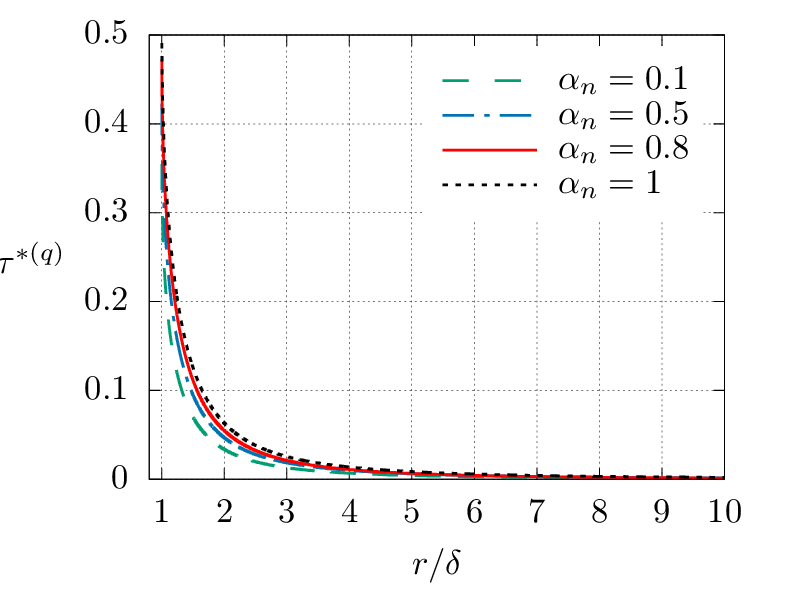}
}
\caption{Flow fields due to the thermodynamic
force $X_q$ for fixed $\alpha_t$=1 and $\delta$=0.1.}
\lae{fig6}
\end{figure}


\begin{figure}
\centering
\subfigure[Radial and polar components of the bulk velocity]{
\includegraphics[scale=0.9]{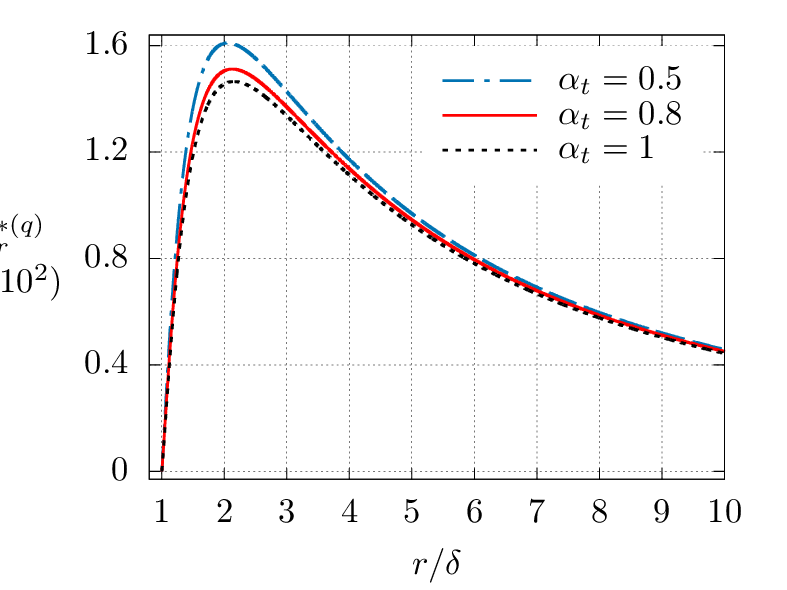}
\includegraphics[scale=0.9]{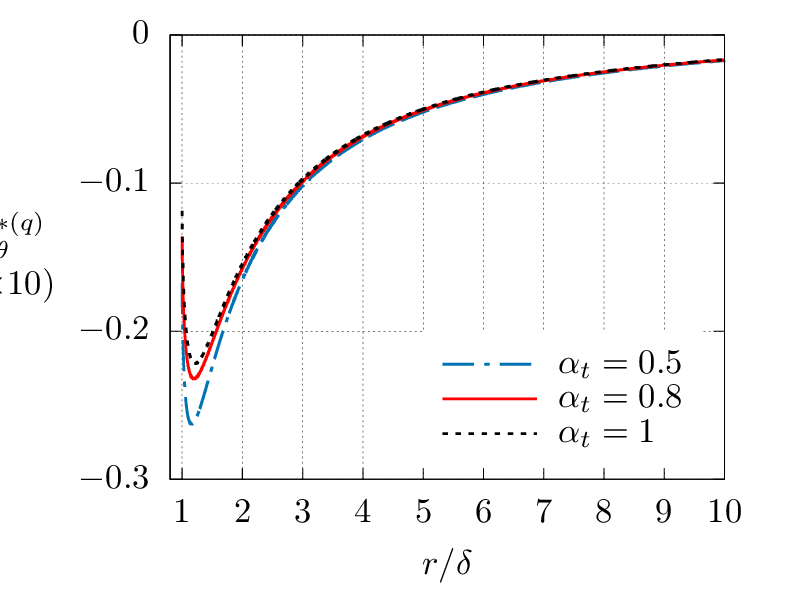}
}
\subfigure[Radial and polar components of the heat flux]{
\includegraphics[scale=0.9]{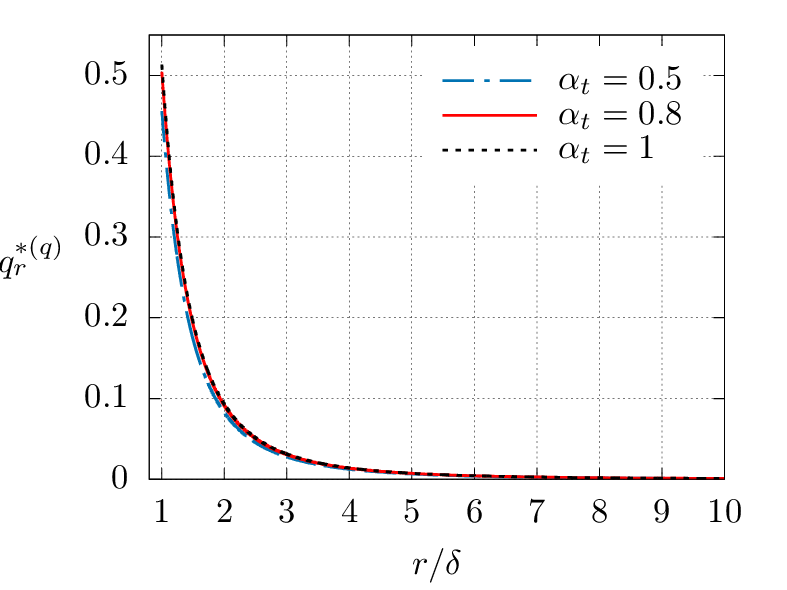}
\includegraphics[scale=0.9]{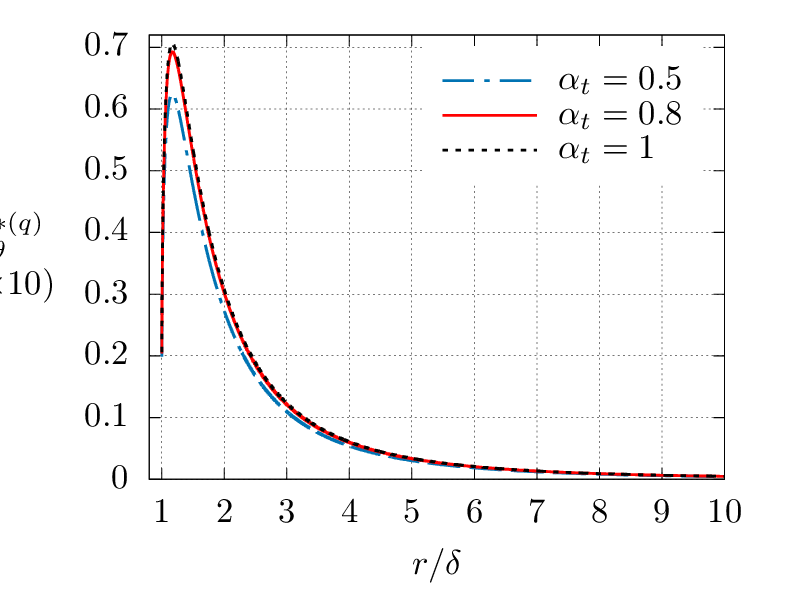}
}
\subfigure[Density and temperature deviations from equilibrium]{
\includegraphics[scale=0.9]{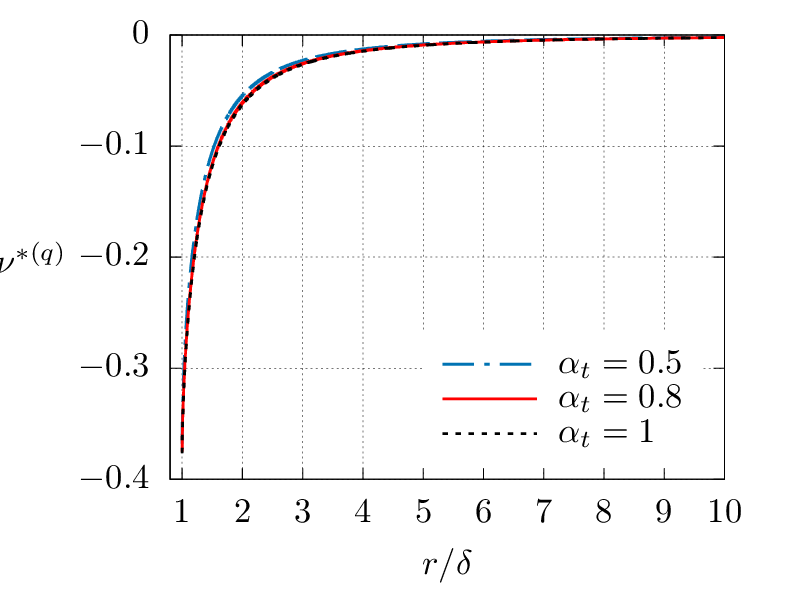}
\includegraphics[scale=0.9]{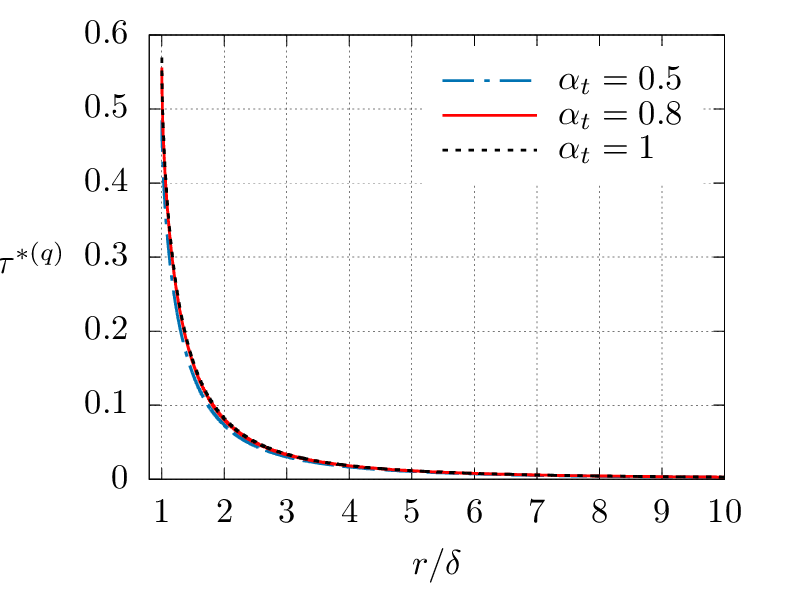}
}
\caption{Flow fields due to the thermodynamic
force $X_q$ for fixed $\alpha_n$=1 and $\delta$=1.}
\lae{fig7}
\end{figure}


\begin{figure}
\centering
\subfigure[Radial and polar components of the bulk velocity]{
\includegraphics[scale=0.9]{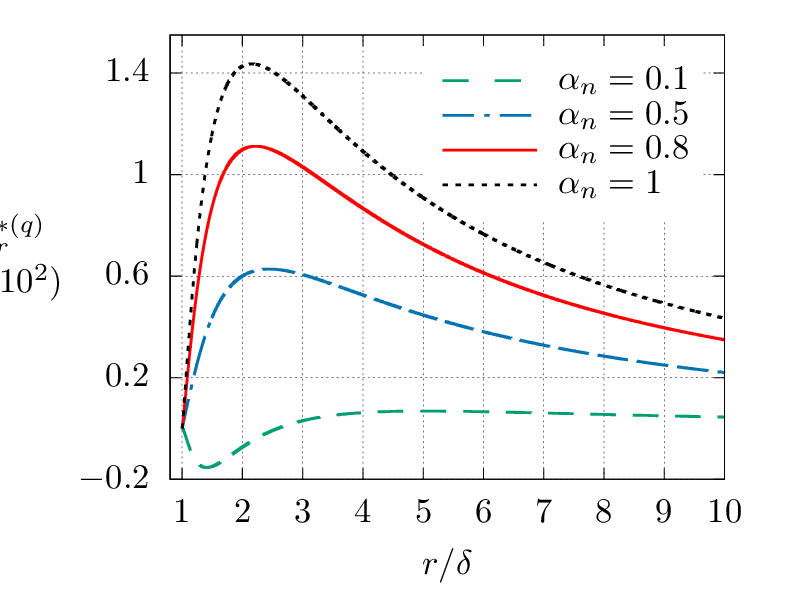}
\includegraphics[scale=0.9]{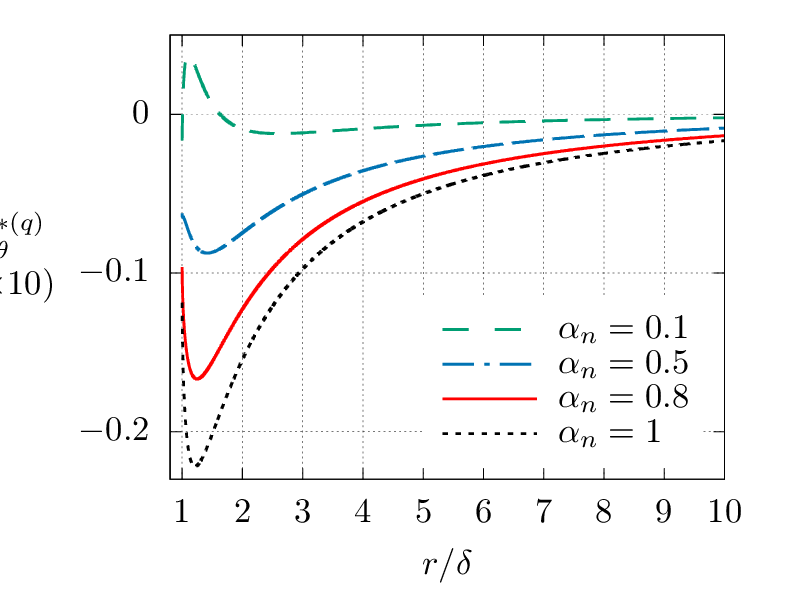}
}
\subfigure[Radial and polar components of the heat flux]{
\includegraphics[scale=0.9]{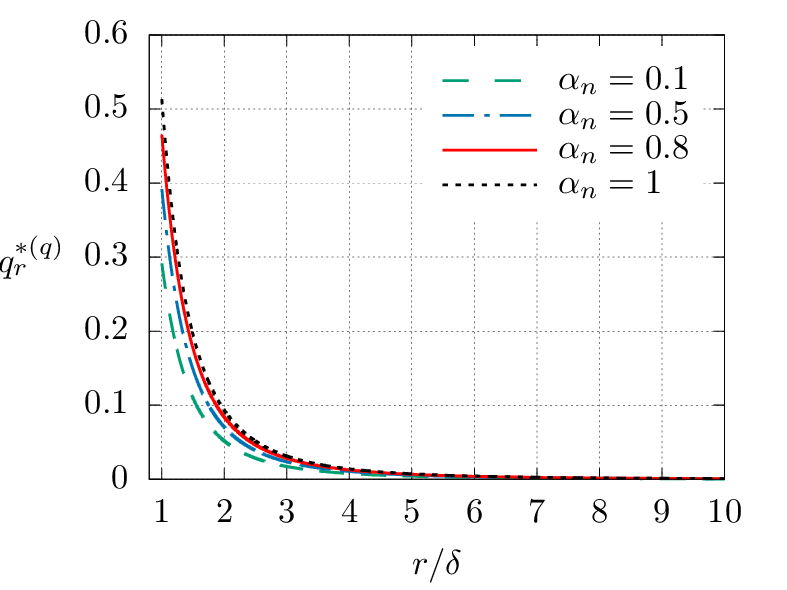}
\includegraphics[scale=0.9]{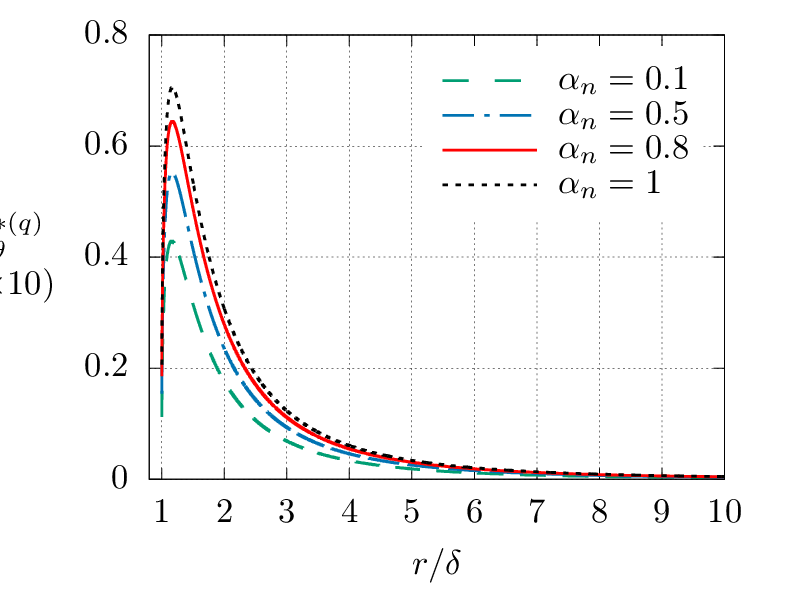}
}
\subfigure[Density and temperature deviations from equilibrium]{
\includegraphics[scale=0.9]{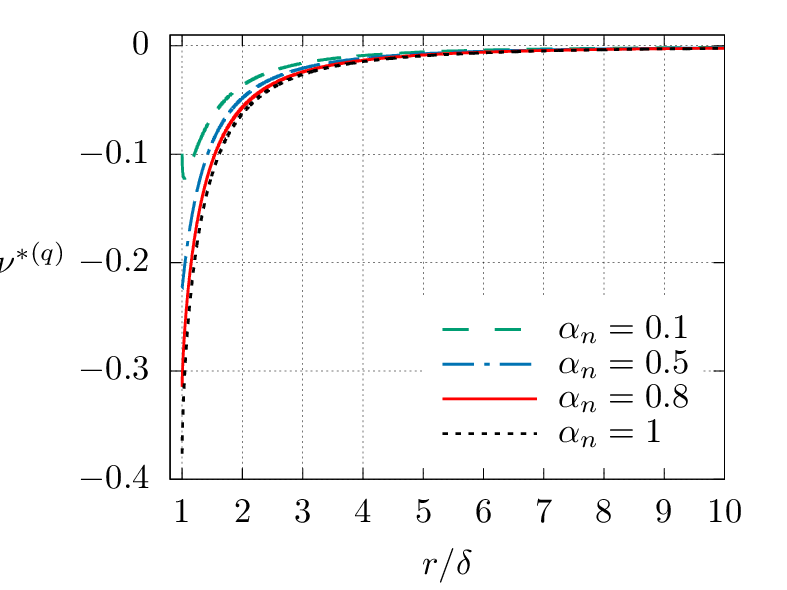}
\includegraphics[scale=0.9]{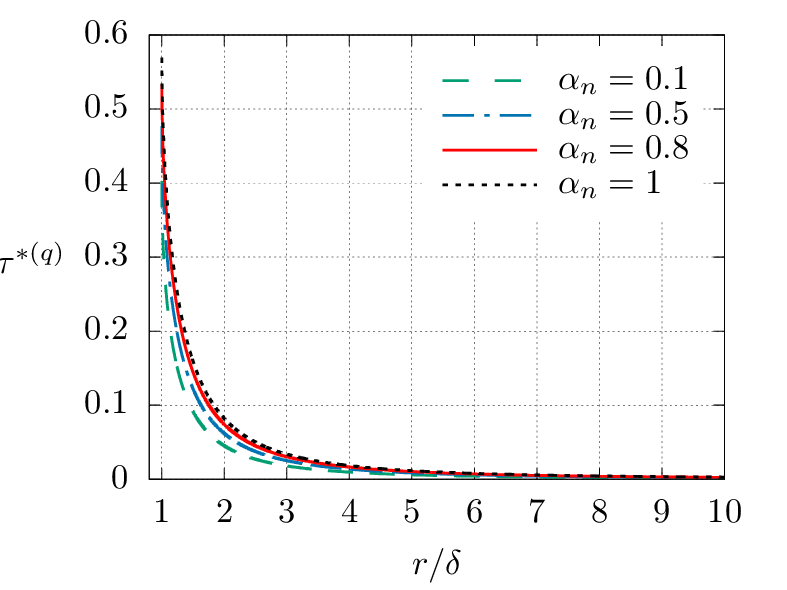}
}
\caption{Flow fields due to the thermodynamic
force $X_q$ for fixed $\alpha_t$=1 and $\delta$=1.}
\lae{fig8}
\end{figure}


\begin{figure}
\centering
\subfigure[Radial and polar components of the bulk velocity]{
\includegraphics[scale=0.9]{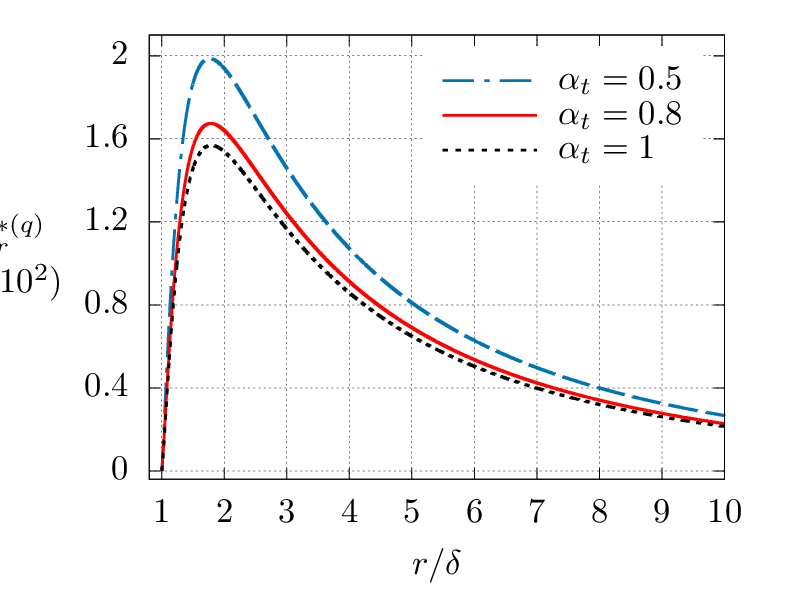}
\includegraphics[scale=0.9]{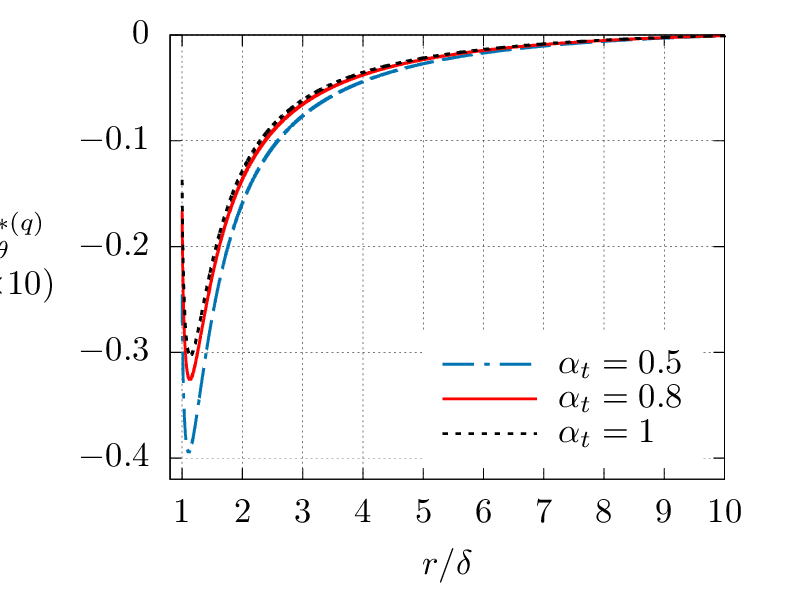}
}
\subfigure[Radial and polar components of the heat flux]{
\includegraphics[scale=0.9]{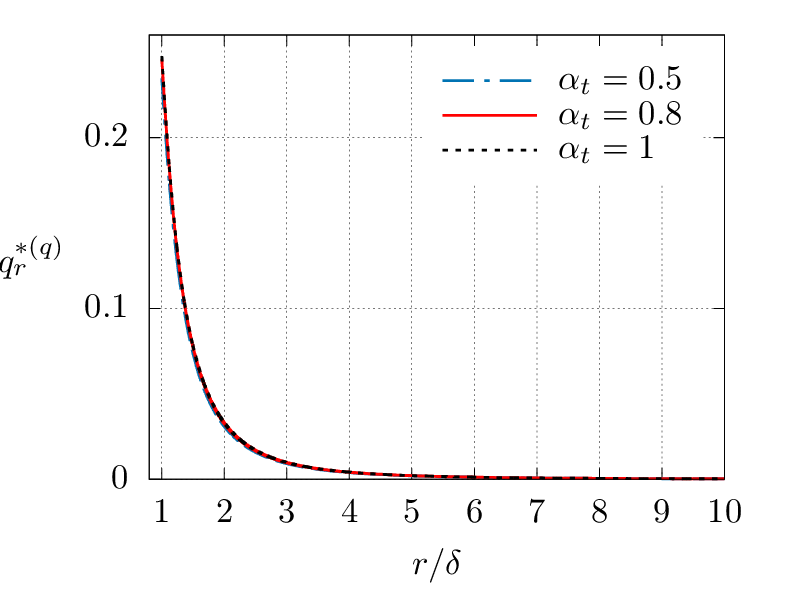}
\includegraphics[scale=0.9]{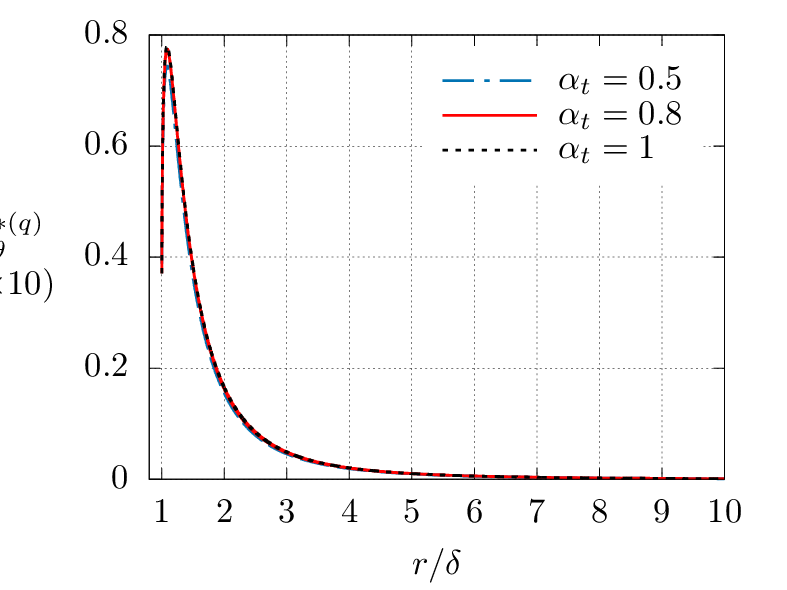}
}
\subfigure[Density and temperature deviations from equilibrium]{
\includegraphics[scale=0.9]{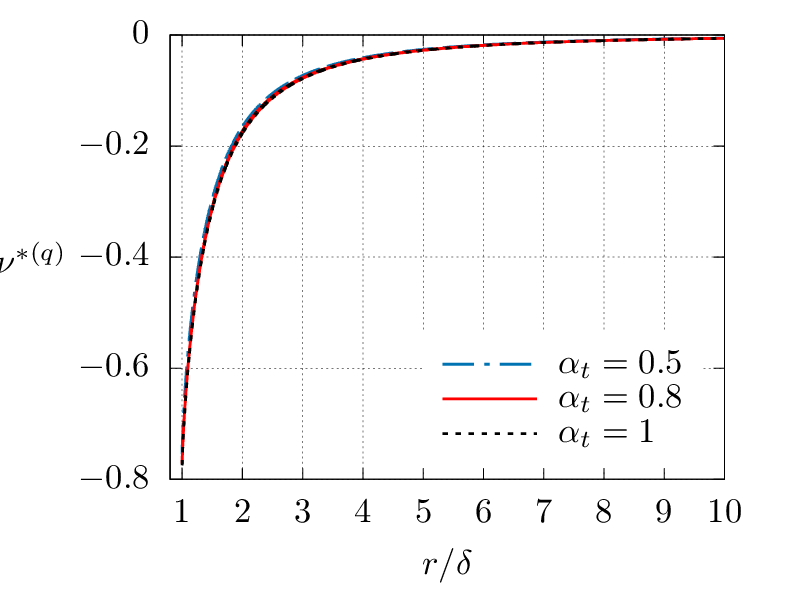}
\includegraphics[scale=0.9]{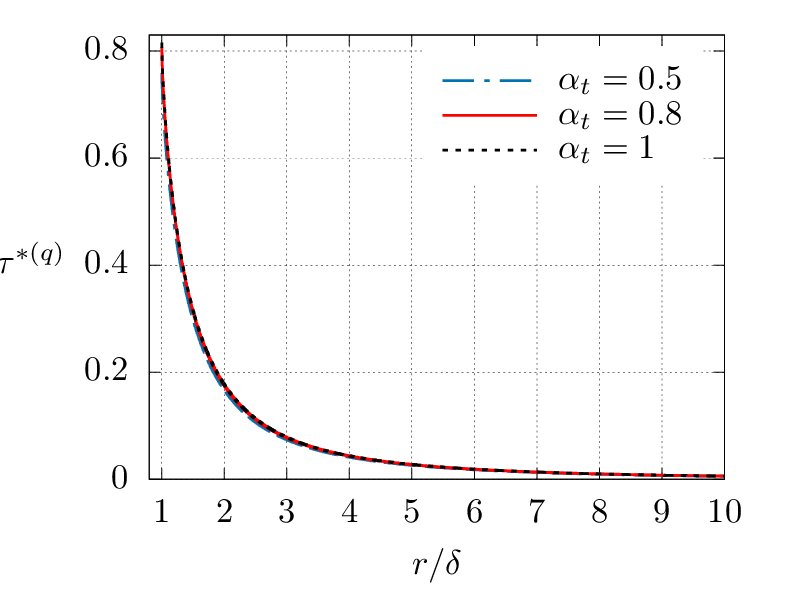}
}
\caption{Flow fields due to the thermodynamic
force $X_q$ for fixed $\alpha_n$=1 and $\delta$=10.}
\lae{fig9}
\end{figure}

\begin{figure}
\centering
\subfigure[Radial and polar components of the bulk velocity]{
\includegraphics[scale=0.9]{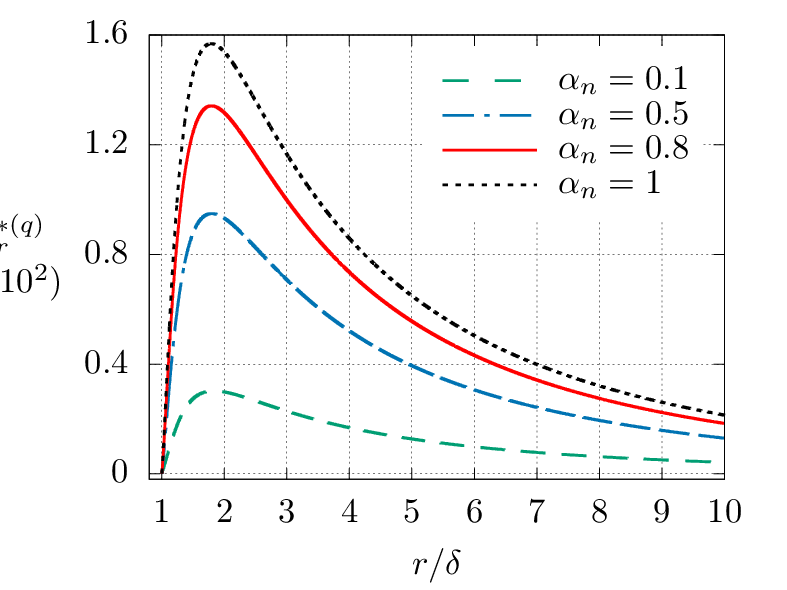}
\includegraphics[scale=0.9]{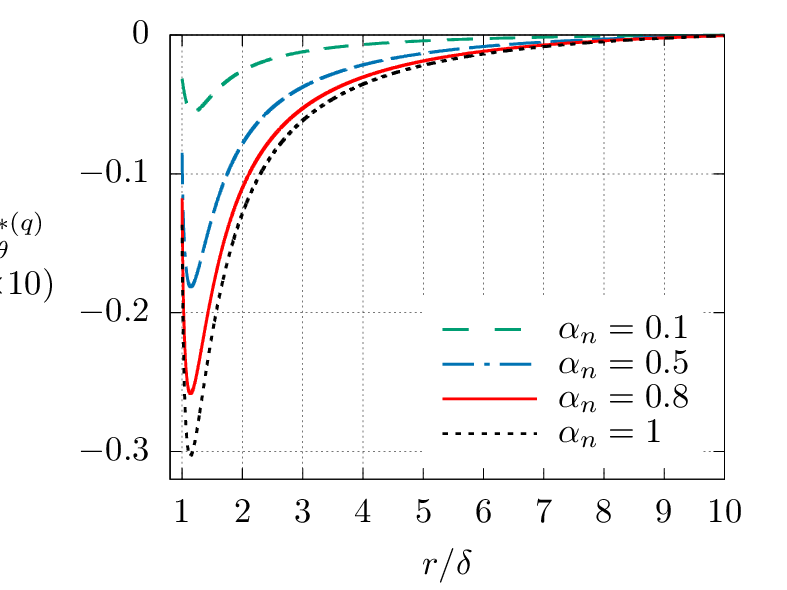}
}
\subfigure[Radial and polar components of the heat flux]{
\includegraphics[scale=0.9]{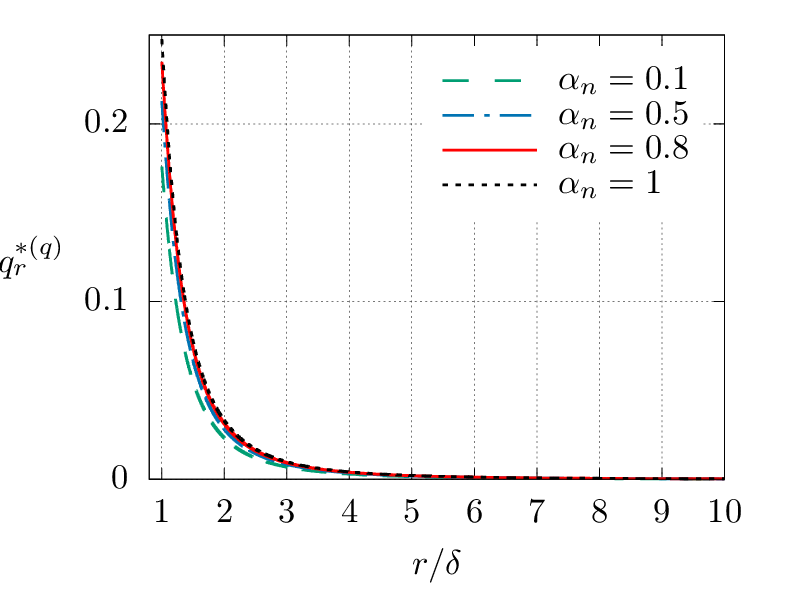}
\includegraphics[scale=0.9]{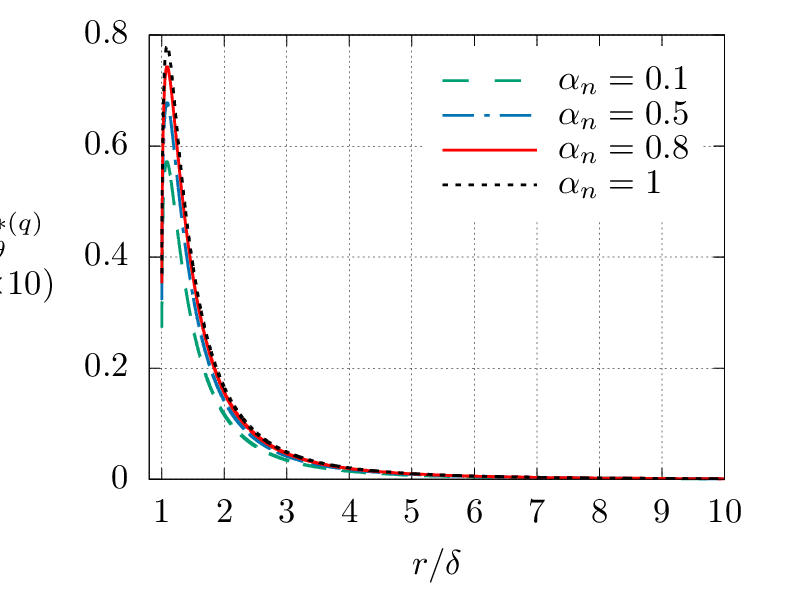}
}
\subfigure[Density and temperature deviations from equilibrium]{
\includegraphics[scale=0.9]{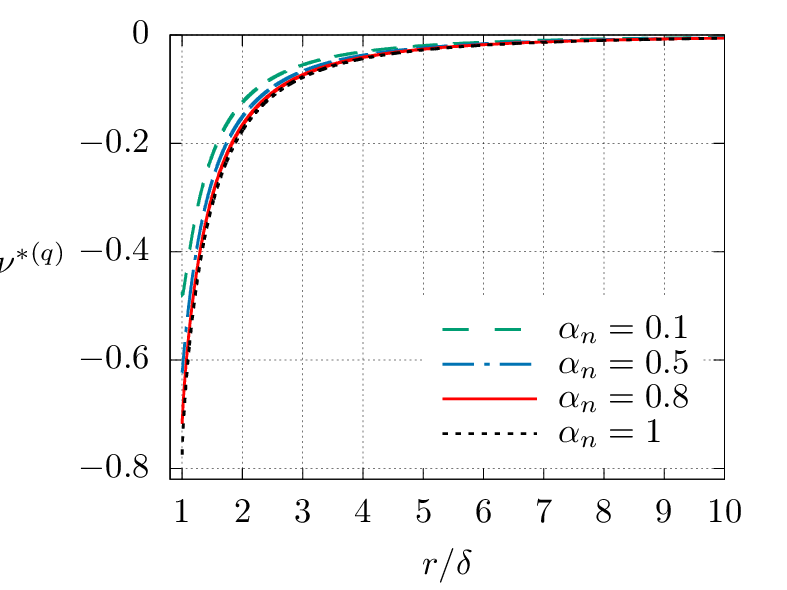}
\includegraphics[scale=0.9]{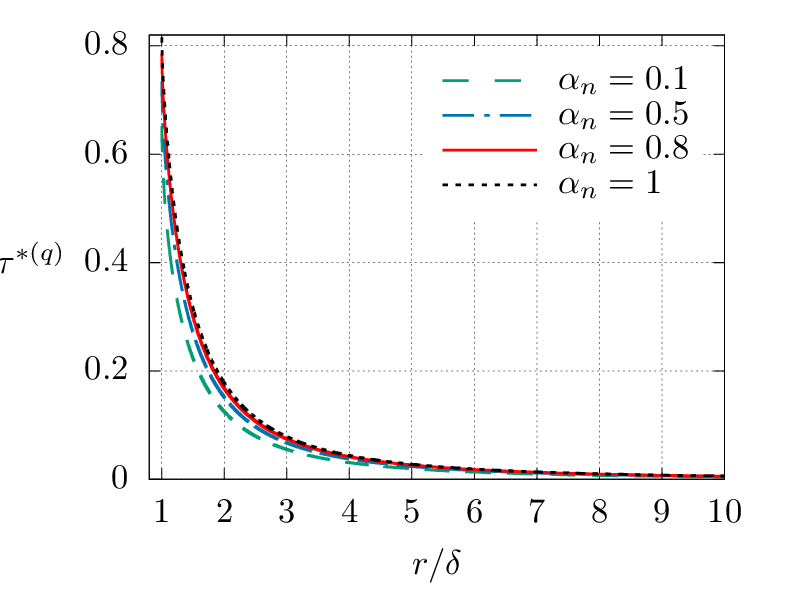}
}
\caption{Flow fields due to the thermodynamic
force $X_q$ for fixed $\alpha_t$=1 and $\delta$=10.}
\lae{fig10}
\end{figure}


\begin{figure}
\centering
\includegraphics[scale=1]{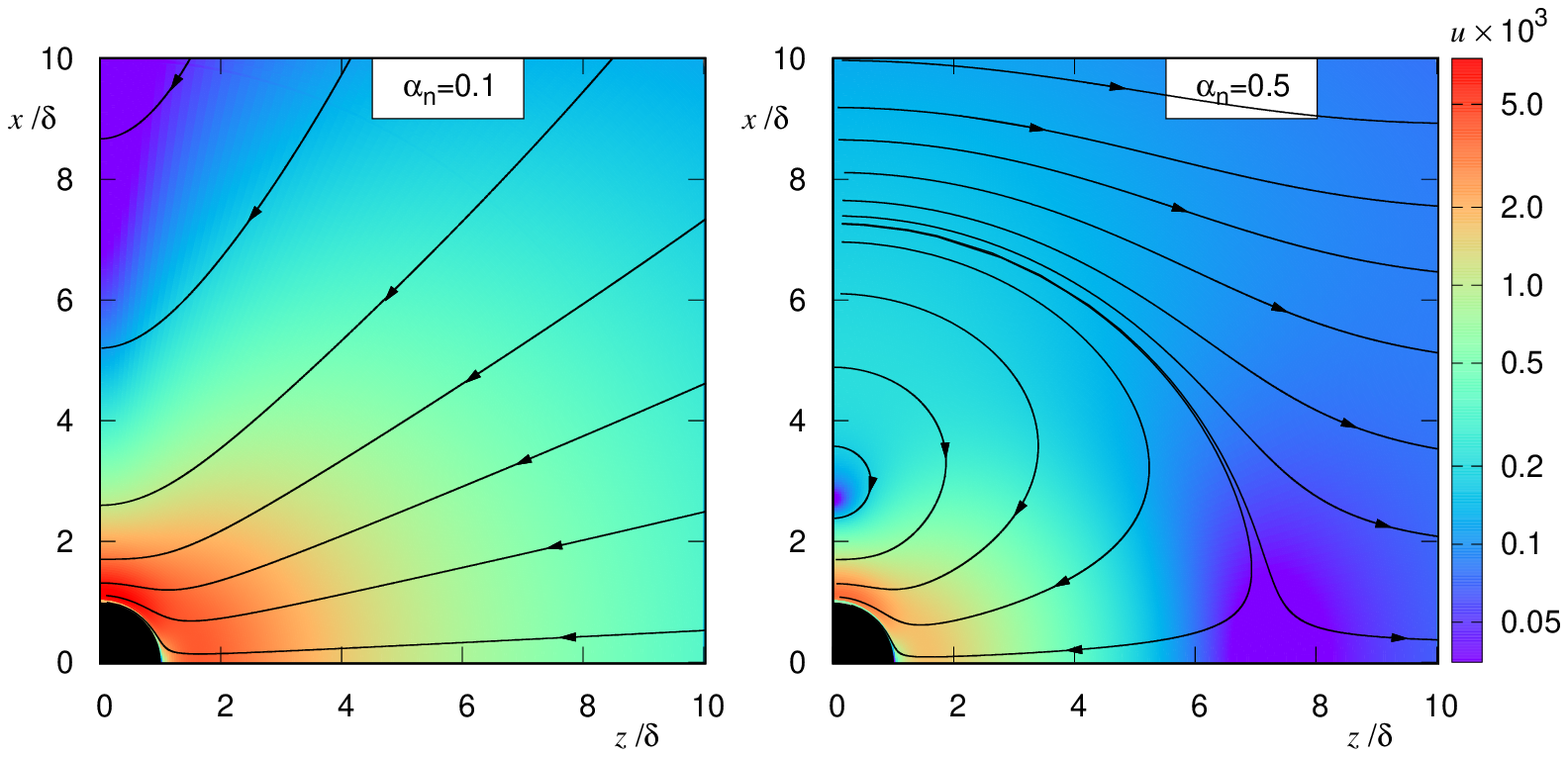}
\caption{Speed contour and velocity streamlines for fixed $\alpha_t=1$ and $\delta$=0.1.}
\lae{mapdel01}
\end{figure}

\begin{figure}
\centering
\includegraphics[scale=1]{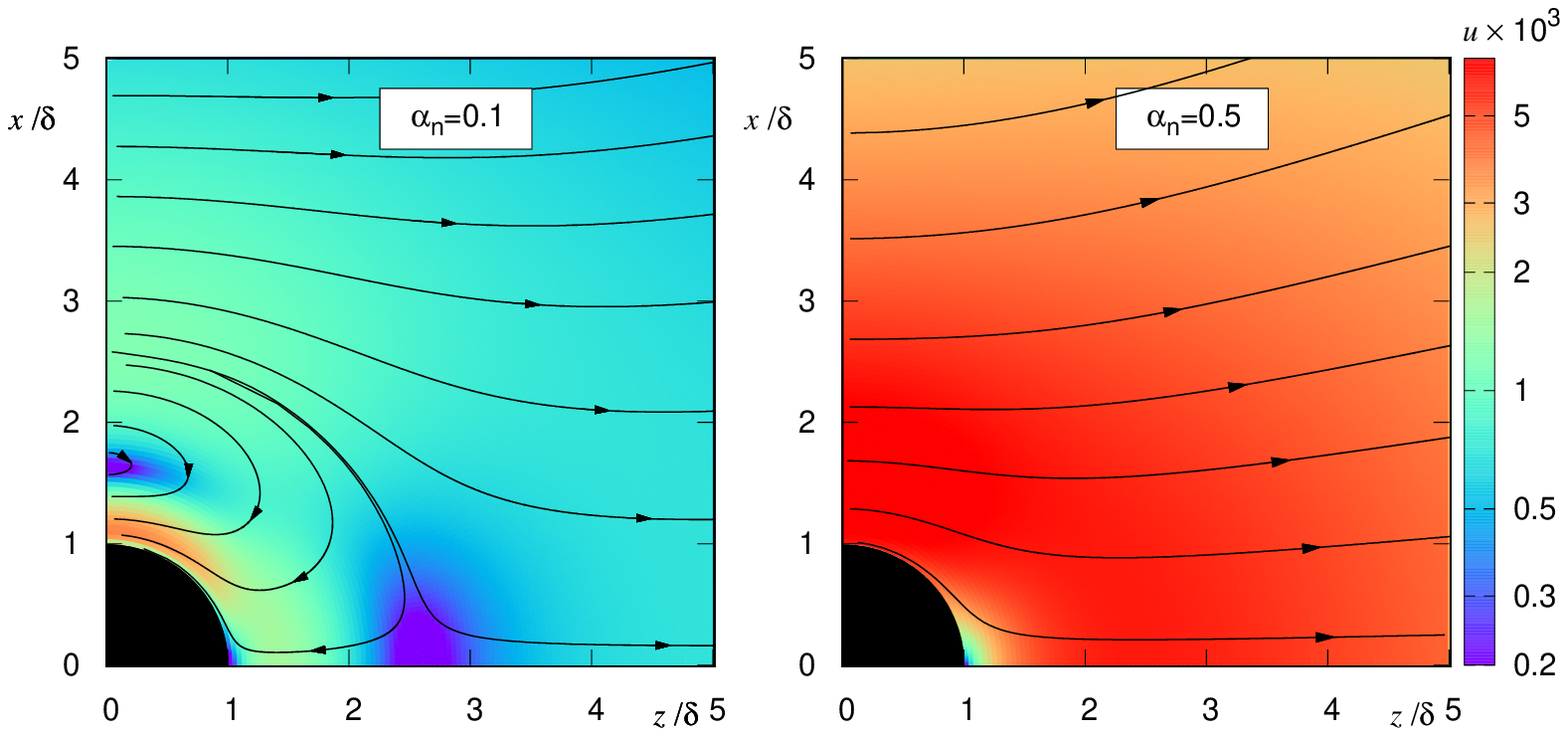}
\caption{Speed contour and velocity streamlines for fixed $\alpha_t=1$ and
$\delta$=1.}
\lae{mapdel1}
\end{figure}

\begin{figure}
\centering
\includegraphics[scale=1]{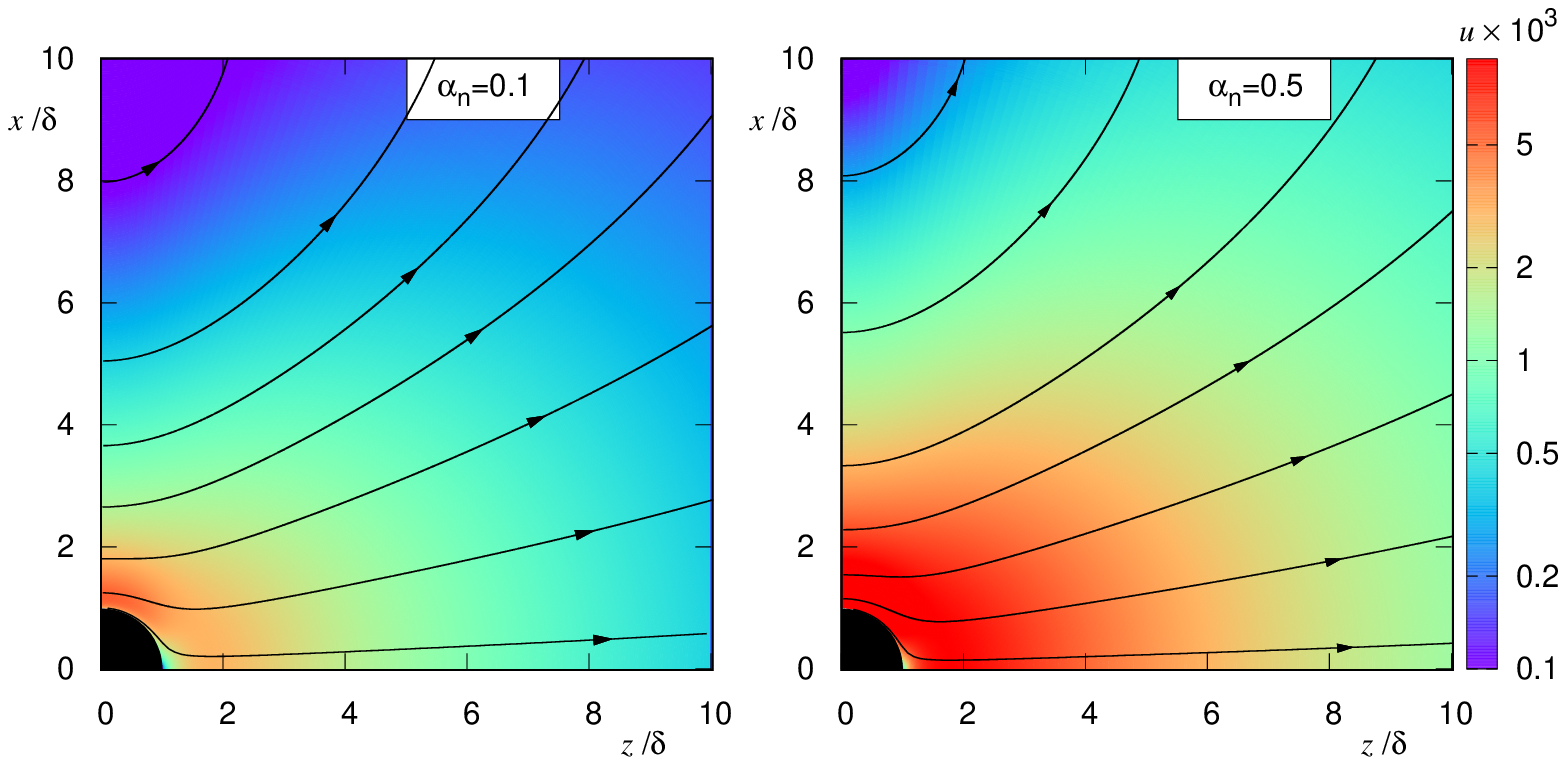}
\caption{Speed contour and velocity streamlines for fixed $\alpha_t=1$ and
$\delta$=10.}
\lae{mapdel10}
\end{figure}

\clearpage

\begin{figure}
\centering
\includegraphics[scale=1.2]{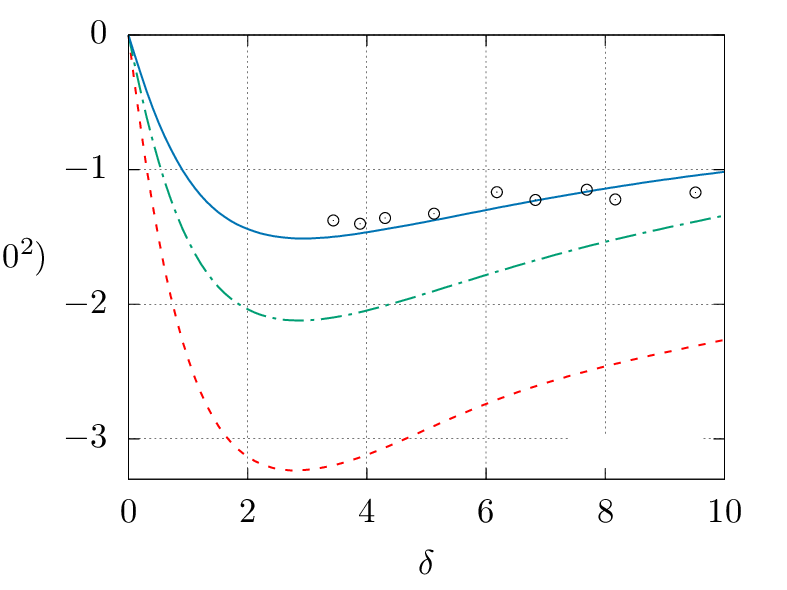}
\caption{Temperature drop for helium gas at a Pyrex glass sphere
($\Lambda$=7.5) as function of the rarefaction
parameter. Red dashed line: $\alpha_n$=1 and $\alpha_t$=1. 
Blue solid line: $\alpha_n$=0.4 and $\alpha_t$=0.9. Green interrupted line: $\alpha_n$=0.6 and
$\alpha_t$=0.9. Symbol $\odot$: experimental data from Ref. \cite{Bak05}.} 
\lae{fig14a}
\end{figure}

\begin{figure}
\centering
\includegraphics[scale=1.2]{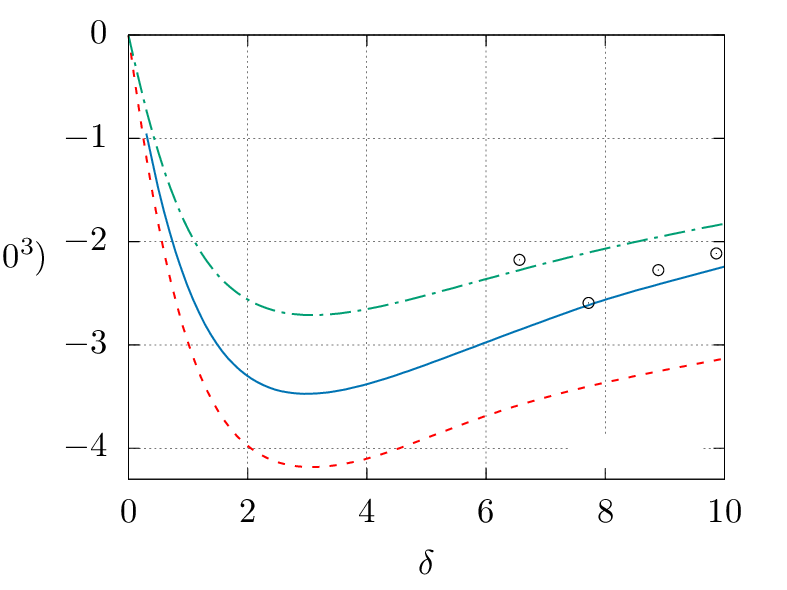}
\caption{Temperature drop for argon gas at a Pyrex glass sphere
($\Lambda$=62.5) as function of the rarefaction
parameter. Red dashed line: $\alpha_n$=1 and $\alpha_t$=1.Blue solid line: $\alpha_n$=0.8 and $\alpha_t$=0.9. 
Green interrupted line: $\alpha_n$=0.6 and
$\alpha_t$=0.9. Symbol $\odot$: experimental data from Ref. \cite{Bak05}.} 
\lae{fig14b}
\end{figure}

\end{document}